\documentclass[groupedaddress,twocolumn,10pt,pra,aps,showpacs,longbibliography,floatfix]{revtex4-2}

\usepackage[dvipsnames]{xcolor}

\definecolor{darkred}{rgb}{0.6,0.05,0.05}
\definecolor{darkgreen}{rgb}{0.05,0.6,0.05}
\definecolor{darkblue}{rgb}{0.05,0.05,0.6}
\usepackage[colorlinks]{hyperref}
\hypersetup{colorlinks,
linkcolor=darkred,
filecolor=darkgreen,
urlcolor=darkblue,
citecolor=darkgreen}

\usepackage{mathtools}
\usepackage{amsmath}
\allowdisplaybreaks
\usepackage{bbold}
\usepackage{amssymb}
\usepackage[normalem]{ulem}
\usepackage{amsthm}
\usepackage{xfrac}
\usepackage{xcolor}
\usepackage{dsfont}
\usepackage{siunitx}
\usepackage{physics}
\usepackage{graphicx}
\usepackage{hyperref}
\usepackage[capitalise]{cleveref}
\usepackage{dcolumn}
\usepackage{bm}
\usepackage{comment}
\usepackage{makecell}
\usepackage{xcolor}
\usepackage[caption=false]{subfig}

\newcommand{\fil}[1]{{\leavevmode\color{blue}{#1}}}

\newcommand{\rma}{{\rm a}}
\newcommand{\rme}{{\rm e}}
\newcommand{\rmd}{{\rm d}}
\newcommand{\rmi}{{\rm i}}
\newcommand{\rmpp}{{\rm p}}
\newcommand{\rmc}{{\rm c}}
\newcommand{\rmf}{{\rm f}}
\newcommand{\bfk}{{\mathbf{k}}}
\newcommand{\bfr}{{\mathbf{r}}}
\newcommand{\bfq}{{\mathbf{q}}}
\newcommand{\bfp}{{\mathbf{p}}}
\newcommand{\bfQ}{{\mathbf{Q}}}

\newcommand{\phantomsubcaptionlabel}[1]{\refstepcounter{subfigure}\label{#1}}

\begin{document}

\preprint{APS/123-QED}

\title{Density Wave Ordering with Disordered Ultracold Fermions in Optical Cavities}

\author{Óscar Rios Alves}
\altaffiliation{These authors contributed equally to this work}
\author{Filippo Ferrari}
\altaffiliation{These authors contributed equally to this work}
\author{Lorenzo Fioroni}
\altaffiliation{Correspondence should be addressed to \fil{filippo.ferrari@epfl.ch} or \fil{lorenzo.fioroni@epfl.ch}}
\author{Alberto Mercurio}
\author{Vincenzo Savona}
\affiliation{Institute of Physics and Center for Quantum Science and Engineering,\\ \'{E}cole Polytechnique F\'{e}d\'{e}rale de Lausanne (EPFL), Lausanne, Switzerland}

\date{\today}

\begin{abstract}
We investigate the interplay between cavity-induced density-wave ordering and controllable disorder in a trapped two-dimensional gas of ultracold fermions.
The atoms are dispersively coupled to an optical cavity and transversely driven by a pump beam, while an additional speckle beam spatially modulates the atom-light coupling through an AC-Stark shift of the atomic transition.
In momentum space, this disorder converts the usual coupling between the cavity mode and a discrete set of density-wave Fourier components into a coupling to a continuum of fermionic density modes, weighted by the spectrum of the speckle pattern.
Using linear response theory, we derive the superradiant threshold and show that the disordered interaction renormalizes the effective light-matter coupling, lowering the critical pump strength on average, with the threshold becoming self-averaging for short speckle correlation lengths.
We complement this analysis with a numerical mean-field treatment that gives access to the intracavity photon number and to the real-space fermion density across the transition.
These results confirm that the disorder shifts the photonic phase boundary and, above threshold, distorts the density-wave crystal by populating Fourier components beyond those selected by the clean cavity geometry.
Our findings identify both the emitted cavity light and \textit{in situ} density images as probes of engineered disorder in fermionic matter coupled to optical cavities.
\end{abstract}

\maketitle

\section{Introduction}

Ultracold quantum gases provide a highly controllable setting to study how collective phases of matter emerge from microscopic Hamiltonians~\cite{bloch_many-body_2008, gross_quantum_2017}.
Fermionic atoms are particularly appealing in this context, because their response is shaped by Fermi statistics, Pauli blocking, and the geometry of the Fermi surface.
At the same time, they remain experimentally challenging~\cite{esslinger_fermi-hubbard_2010, hart_observation_2015, mazurenko_cold-atom_2017}: ordered phases can be fragile, their signatures may be encoded in nonlocal correlations, and the relevant many-body dynamics often occurs in regimes where controlled theoretical descriptions are difficult.
Developing platforms in which fermionic order can be created, tuned, and read out in real time~\cite{ polkovnikov_colloquium_2011, bloch_quantum_2012} is therefore a central goal of quantum simulation.

Optical cavities offer a powerful route towards this goal~\cite{ritsch_cold_2013, mivehvar_cavity_2021}.
When atoms are placed inside a driven cavity, photons scattered between the pump and the cavity mediate interactions that are long-ranged on the scale of the atomic cloud.
Above a critical pump strength, the atoms can self-organize into a density-wave-ordered (DWO) state accompanied by a macroscopic coherent cavity field: a superradiant phase transition~\cite{domokos_collective_2002, black_observation_2003, baumann_dicke_2010, nagy_dicke-model_2010, leonard_supersolid_2017, kollar_supermode-density-wave-polariton_2017, kroeze_spinor_2018}.
This transition is intrinsically collective, since the density modulation of the atoms enhances the cavity field, while the cavity field deepens the optical potential that favors the modulation.
It is also directly observable~\cite{mivehvar_cavity_2021}, as the onset of order appears both in the light leaking out of the cavity~\cite{baumann_dicke_2010} and in the momentum-space density profile of the atoms~\cite{baumann_dicke_2010}.
For fermions, the instability is controlled by the density response of the Fermi gas, making cavity QED a natural arena for probing how fermionic susceptibilities shape ordered states~\cite{keeling_fermionic_2014, piazza_umklapp_2014, chen_superradiance_2014, zhang_observation_2021, helson_density-wave_2023, zwettler_cavity-mediated_2025, buhler_microscopy_2026, orsi2026fermipressureassistedcavitysuperradiancemesoscopic}.

A second major direction in quantum simulation with ultracold atoms is the controlled introduction of disorder.
Speckle fields and related optical techniques make it possible to engineer spatial randomness with tunable strength and correlation length, enabling studies of localization~\cite{clement_suppression_2005, billy_direct_2008, roati_anderson_2008, kondov_three-dimensional_2011, jendrzejewski_three-dimensional_2012, schreiber_observation_2015, choi_exploring_2016, luschen_observation_2017, lukin_probing_2019}, glassiness~\cite{gopalakrishnan_frustration_2011, marsh_enhancing_2021, marsh_entanglement_2024, kroeze_directly_2025, marsh_multimode_2025}, and random quantum many-body systems~\cite{lippe_experimental_2021, periwal_programmable_2021, sauerwein_engineering_2023} under conditions that are difficult to access in solid-state materials.
In many settings disorder is introduced as an external potential acting directly on the particles~\cite{clement_suppression_2005} or through the use of a multimode cavity~\cite{kroeze_directly_2025}.
Here we consider a different possibility: disorder in the light-matter interaction itself.
An additional optical beam with a spatially random intensity profile AC-Stark shifts the atomic transition frequency, producing a position-dependent atom-light coupling while leaving the cavity geometry and pump configuration otherwise intact~\cite{orsi_cavity_2024}.
This realizes a controllable disordered interaction between the fermionic density and the cavity field, that can potentially lead to exotic phases of matter~\cite{uhrich_cavity_2023, baumgartner_quantum_2024, baumgartner_quantum_2025, solis_single-particle_2026}.

The central question addressed in this work is how such engineered disorder modifies cavity-induced density-wave ordering~\cite{helson_density-wave_2023}.
In a clean transversely pumped cavity, the light field couples only to density Fourier components at the momenta fixed by the pump and cavity wavevectors, $\bfQ=\pm(\bfk_\rmc\pm\bfk_\rmpp)$.
The speckle modulation changes this structure qualitatively: in Fourier space the disorder convolves the clean coupling with the speckle spectrum, so that the cavity field couples to a continuum of fermionic density modes.
This additional momentum mixing can enhance the effective fermion-cavity response and thereby favor the onset of superradiance, but it can also imprint spatial randomness on the ordered state once the transition has occurred.

We analyze these effects using two complementary approaches.
First, we derive the DWO phase boundary from a linear response treatment of the coupled cavity and fermion equations of motion.
This yields an analytical expression for the critical pump strength in terms of the Lindhard susceptibility and the Fourier components of the speckle-modulated coupling.
It also allows us to study the disorder-averaged threshold and its fluctuations as a function of the speckle correlation length $\xi$.
In the short-correlation-length limit, the system becomes self-averaging and the transition is shifted in a universal way that can be interpreted as a renormalization of the clean light-matter coupling.
Second, we solve mean-field steady-state equations for the coherent cavity amplitude and the single-particle fermionic density matrix, which gives access to both photonic observables and the real-space density profile across the transition.

We find that controllable disorder leaves clear signatures in both sectors.
On the photonic side, the critical pump strength is reduced relative to the clean system for the parameters considered, and the dependence of the threshold on $\xi$ reflects the crossover between self-averaging disorder and an effectively clean coupling.
On the atomic side, the clean checkerboard-like density wave is distorted by the speckle-induced coupling to additional Fourier components.
This distortion is weak when the speckle varies on scales much shorter than the cloud size, and it also disappears when the disorder is effectively constant over the cloud; it is most visible at intermediate correlation lengths, where the speckle grains are resolved by the ordered fermionic density.
The emitted cavity light therefore provides a sensitive global probe of the disorder-shifted transition, while time-of-flight measurements of the fermion momentum distribution~\cite{baumann_dicke_2010} or direct \emph{in situ} density imaging~\cite{buhler_microscopy_2026} reveal how the same disorder reshapes the microscopic DWO pattern.

The work is organized as follows.
In \cref{sec:cavity_qed}, we introduce the cavity-QED model with a speckle-modulated light-matter coupling and describe the analytical linear response and numerical mean-field methods used to determine the DWO transition.
In \cref{sec:results}, we present the main results, focusing first on photonic observables and the disorder-dependent superradiant threshold, and then on the distortion of the fermionic density-wave pattern.
In \cref{sec:conclusion}, we summarize our conclusions and outline future directions.
The appendices provide details on the derivation of the threshold, its disorder average and variance, the trap averaging, the numerical construction of the speckle pattern, and additional analyses of the $\mathbb{Z}_2$ symmetry and the dependence of the threshold on the Fermi wave vector $k_F$, the fermion number and the atom-drive detuning.

\textbf{Note added:} While completing this manuscript, we became aware of a related complementary work titled \emph{``Disorder-Induced Enhancement of Fermionic Superradiance"}, by David Pascual Solis \emph{et. al.}. This will appear on the preprint server simultaneously with this work, and presents a study of superradiance in fermionic atoms subject to Gaussian random disorder in the light-matter couplings.

\section{Cavity QED with ultracold fermions}\label{sec:cavity_qed}

\begin{figure}[t]
    \centering
    \includegraphics[]{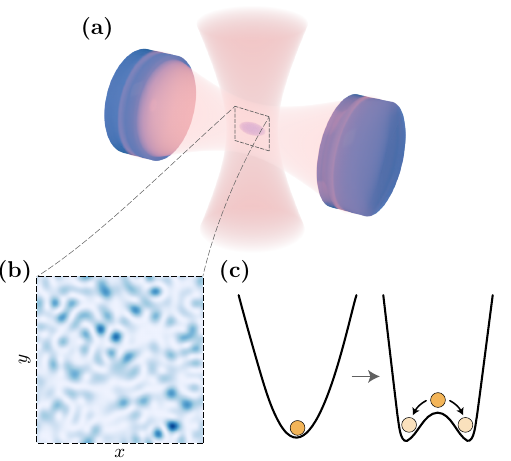}
    \caption{
    Sketch of the cavity setup. (a) An ultracold gas of fermions is harmonically trapped in an optical cavity and driven by a pump laser orthogonal to the cavity axis. (b) Superimposed speckle potential. (c) Sketch of the free energy and the spontaneous symmetry breaking associated with the DWO phase transition~\cite{helson_density-wave_2023}.
    }
    \label{fig:sketch}
    {\phantomsubcaptionlabel{fig:sketch platform}}
    {\phantomsubcaptionlabel{fig:sketch potential}}
    {\phantomsubcaptionlabel{fig:sketch free energy}}

\end{figure}

\subsection{Model}\label{sec:model}

We consider a trapped two-dimensional gas of fermionic atoms dispersively coupled to an optical cavity with wave vector $\bfk_\rmc$.
The cavity is driven by a transverse pump with wave vector $\bfk_\rmpp$.
The interference between the pump and cavity fields creates an optical potential and mediates long-range interactions among the fermions. Depending on the detuning from the cavity resonance, these interactions can be interpreted as being mediated by virtual or real photons.
In the frame rotating at the pump frequency, the Hamiltonian reads
\begin{equation}
\label{eq:cQED_Hamiltonian}
    \hat{H} =\hat{H}_{\rmf} + \frac{\Delta_{\rmc\rmd}}{2}\left(\hat{x}^2 + \hat{p}^2\right) + \hat{H}_{\rm int},
\end{equation}
where $\hat{H}_\rmf$ describes non-interacting fermions confined by a harmonic trap, $\Delta_{\rmc\rmd}=\omega_\rmc-\omega_\rmd$ is the cavity-drive detuning, $\hat{x}$ and $\hat{p}$ are cavity-field quadratures, and $\hat{H}_{\rm int}$ is the light-matter interaction Hamiltonian.
The entire setup is sketched in \cref{fig:sketch platform}.
We introduce disorder in the interaction by adding a beam that AC-Stark shifts the atomic transition frequency and has a spatially random intensity $I(\bfr)$ with speckle correlation length $\xi$; the numerical construction of $I(\bfr)$ is described in Appendix~\ref{app:appendix_A}, while an example of a typical speckle intensity is reported in \cref{fig:sketch potential}.
The resulting interaction Hamiltonian is
\begin{equation}\label{eq:int_hamiltonian}
    \hat{H}_{\rm int} = \frac{\Omega_\rmd\Omega}{\sqrt{2}\Delta_{\rma\rmd}}\hat{x}\int \rmd\bfr\,\frac{\cos(\bfk_\rmpp\cdot\bfr)\cos(\bfk_\rmc\cdot\bfr)\,\hat{n}(\bfr)}{1 + I(\bfr)/\langle I\rangle},
\end{equation}
where $\Delta_{\rma\rmd}=\omega_\rmd-\omega_\rma$ is the atom-drive detuning, $\Omega$ and $\Omega_\rmd$ are the cavity and drive Rabi frequencies, respectively, and $\hat{n}(\bfr)=\hat{\Psi}^\dagger(\bfr)\hat{\Psi}(\bfr)$ is the fermionic density.
Throughout the manuscript we compare the disordered interaction in \cref{eq:int_hamiltonian} with the clean case, obtained as the limit where $I(\bfr)$ is constant over the cloud.
Notice that here $I(\bfr)$ is renormalized by its spatial average $\langle I(\bfr) \rangle$, which encodes the effective power of the speckle beam. Different choices of the power can lead to weaker disorder strengths. As we point out later, this renormalization leaves the spatial disorder as the only source of randomness in the system, ruling out the random mean shift in $\Delta_{\rmd\rma}$ as the speckle correlation length diverges.
Cavity dissipation is included through the Lindblad jump operator $\hat{L}=\sqrt{\kappa}\hat{a}$, with $\kappa$ the single-photon loss rate.

\begin{figure*}[t]
    \centering
    \includegraphics[width=\linewidth]{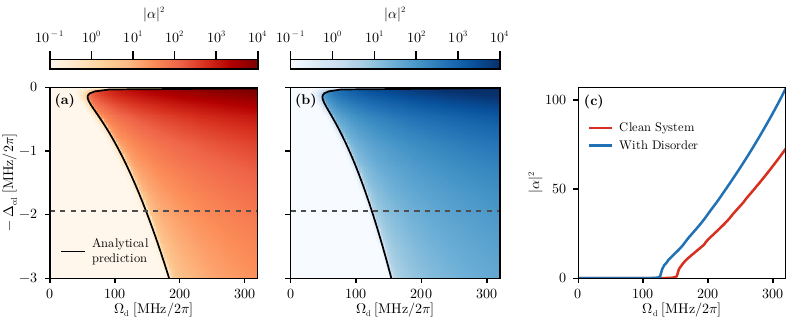}
    \caption{Phase diagrams of the cavity superradiance transition in the $\Omega_\rmd-\Delta_{\rmc\rmd}$ plane. (a) Phase diagram in the absence of disorder. (b) Phase diagram in the presence of disorder, for a single disorder realization with $\xi = 0.1 \mu\mathrm{m}$. The color scale shows the photon number $|\alpha|^2$, obtained from the fixed points of the mean-field equations of motion. The numerical results are compared with the analytical phase boundary derived from linear response theory, as in  \cref{eq:phase_boundary}. (c) Photon number as a function of $\Omega_\rmd$ at fixed $\Delta_{\rmc\rmd}$, indicated by the dashed lines in panels (a) and (b), comparing the clean and disordered cases.
    Parameters are fixed to $\Delta_{\rmd\rma}/2\pi = 1\,\textrm{GHz}$, $\Omega/2\pi=4\,\textrm{MHz}$, $\kappa/2\pi=300\,\textrm{kHz}$, $\lambda_p=\lambda_c=671\,\textrm{nm}$, $x_0=y_0=0.27\,\mu\textrm{m}$.}
    \label{fig:phase_diagrams}
    \phantomsubcaptionlabel{fig:phase_diagrams clean}
    \phantomsubcaptionlabel{fig:phase_diagrams disordered}
    \phantomsubcaptionlabel{fig:phase_diagrams photons}
\end{figure*}

\begin{figure}[h]
    \centering
    \includegraphics[width=\linewidth]{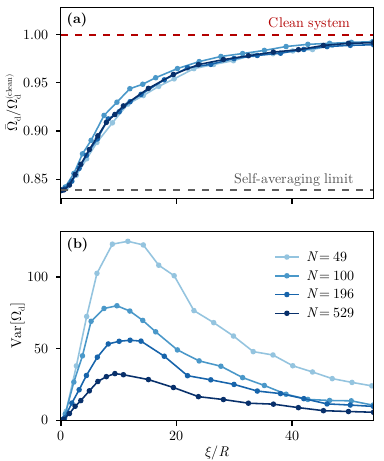}
    \caption{(a) Superradiant threshold, computed from  \cref{eq:phase_boundary} and averaged over 1000 disorder realizations, as a function of the speckle correlation length $\xi$ and for different fermion numbers. The critical drive strength is shown relative to its value in the clean system, and the correlation length is expressed in units of the fermion-cloud radius $R$ determined within the local density approximation. 
    The gray (red) dashed line shows the $\xi\to0$ ($\xi \to \infty$) prediction of  \cref{eq:phase_boundary_disorder_avg}, with $\mu\simeq0.6$ ($\mu \to 0.5$). (b) Variance of the superradiant threshold across disorder realizations as a function of $\xi$.
    Parameters as in \cref{fig:phase_diagrams}.}
    \label{fig:disorder_avg_var}
    \phantomsubcaptionlabel{fig:disorder_avg_var threshold}
    \phantomsubcaptionlabel{fig:disorder_avg_var variance}
\end{figure}

\begin{figure*}[t]
    \centering
    \includegraphics[width=\linewidth]{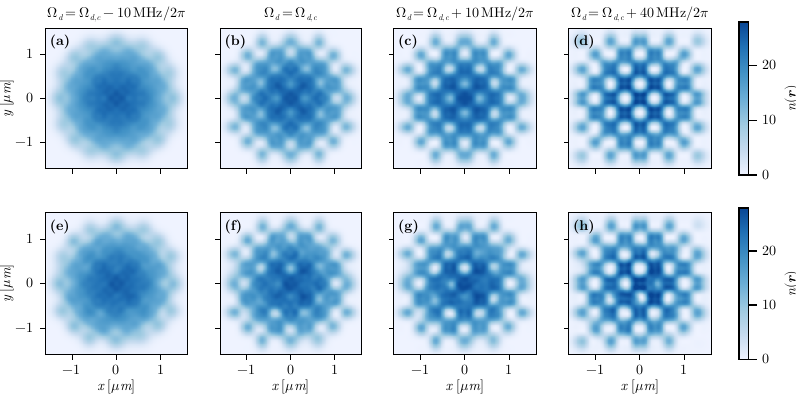}
    \caption{Formation of the DWO crystal in the fermion density across the transition. The density $n(\bfr)$, obtained from the steady state of the mean-field equations, is shown for different values of $\Omega_\rmd$ expressed relative to the analytical critical value $\Omega_{\rmd,\rmc}$. Upper panels: clean system. Lower panels: disordered system with $\xi=\lambda_{\rm p}$. Parameters as in \cref{fig:phase_diagrams}.}
    \label{fig:DWO_transition}
    \phantomsubcaptionlabel{fig:DWO_transition a}
    \phantomsubcaptionlabel{fig:DWO_transition b}
    \phantomsubcaptionlabel{fig:DWO_transition c}
    \phantomsubcaptionlabel{fig:DWO_transition d}
    \phantomsubcaptionlabel{fig:DWO_transition e}
    \phantomsubcaptionlabel{fig:DWO_transition f}
    \phantomsubcaptionlabel{fig:DWO_transition g}
    \phantomsubcaptionlabel{fig:DWO_transition h}
\end{figure*}

\begin{figure}[t]
    \centering
    \includegraphics[width=\linewidth]{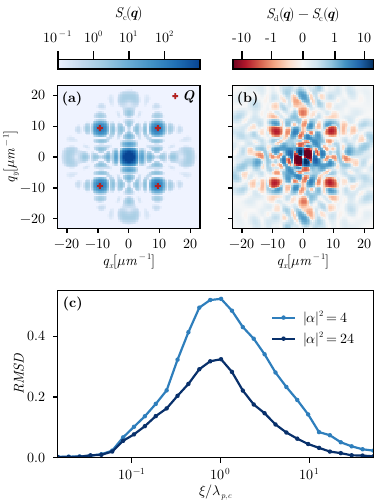}
    \caption{(a) Structure factor $S_\rmc(\bfq)$ of the fermion density above the phase transition for a clean system, at a drive strength corresponding to an arbitrary cavity photon number $|\alpha|^2=24$. (b) Difference between the structure factor of a disordered system $S_\rmd(\bfq)$, with correlation length $\xi=\lambda_p$, and that of a clean system, evaluated at the same cavity photon number. (c) Normalized $L^2$ distance between the structure factor of the disordered system and that of the clean system as a function of the correlation length and for different photon numbers. Each curve is averaged over 100 disorder realizations and $\xi$ is expressed in units of $\lambda_p\approx\lambda_c$. Parameters as in \cref{fig:phase_diagrams}.}
    \label{fig:structure_factor}

    \phantomsubcaptionlabel{fig:structure_factor clean}
    \phantomsubcaptionlabel{fig:structure_factor disordered}
    \phantomsubcaptionlabel{fig:structure_factor rmsd}
\end{figure}

To clarify the effect of the spatially disordered coupling, we write \cref{eq:int_hamiltonian} in momentum space.
Let $\mathcal{D}(\bfq) = \mathcal{F}\qty[1/\qty(1 + I/\langle I \rangle)]$ denote the Fourier transform of the speckle-dependent factor and let $\hat{\rho}(\bfq)$ denote the Fourier component of the fermionic density.
The interaction Hamiltonian becomes
\begin{equation}
\label{eq:int_Hamiltonian_fourier_space}
    \hat{H}_{\rm int} = \frac{\Omega_\rmd\Omega}{\sqrt{32}\Delta_{\rma\rmd}}\hat{x}\sum_{\bfQ} \int\rmd\bfq\,\mathcal{D}(\bfQ-\bfq) \hat{\rho}(\bfq),
\end{equation}
where the sum runs over $\bfQ = \pm(\bfk_\rmc\pm \bfk_\rmpp)$.
In the clean case, $\mathcal{D}(\bfQ - \bfq) \propto \delta(\bfQ - \bfq)$, so only the density components $\hat \rho(\bfQ)$ associated with the usual DWO state couple to the cavity field.
In the presence of speckle disorder, all density Fourier components can couple to $\hat x$, with spectral weight $\mathcal D(\bfQ - \bfq)$.
Thus, the disordered interaction mixes the cavity-selected momenta with a continuum of fermionic density modes.

\subsection{Methods}\label{sec:methods}

We first derive the analytical phase boundary of the DWO transition, whose free energy characterization is sketched in \cref{fig:sketch free energy}, by linearizing the cavity and fermion equations of motion within linear response theory~\cite{marijanovic_quench_2026}.
The equations for the cavity field in the frequency domain are
\begin{equation}\label{eqs:cavity_eom}
\begin{aligned}
    \rmi\omega x &= \Delta_{\rmc\rmd}{p}-\frac\kappa2 x, \\
    \rmi\omega p &= -\Delta_{\rmc\rmd}x-\frac\kappa2 p - \eta\sum_{\bfQ} \int \rmd\bfq\,\mathcal{D}(\bfQ-\bfq)\rho(\bfq),
\end{aligned}
\end{equation}
where $\eta = \Omega_\rmd \Omega /\sqrt{32}\Delta_{\rma\rmd}$.
The corresponding linearized density response is
\begin{equation}\label{eq:fermion_eom}
    \rho(\bfq) =  \eta\,x\sum_{\bfQ} \mathcal{D}(\bfQ - \bfq) \, \chi(\bfq, \omega),
\end{equation}
where $\chi(\bfq, \omega)$ is the Lindhard function of noninteracting fermions~\cite{giuliani_quantum_2005}.
The onset of DWO is determined by the appearance of solutions of \cref{eqs:cavity_eom,eq:fermion_eom} for which the intracavity field grows exponentially.
As derived in Appendix~\ref{app:appendix_A}, the resulting phase boundary is
\begin{equation}
\begin{split}
\label{eq:phase_boundary}
    \Delta_{\rmc\rmd}+\frac{(\kappa/2)^2}{\Delta_{\rmc\rmd}}&=\\
    -\frac{\Omega_\rmd^2\Omega^2}{32\Delta_{\rma\rmd}^2}\sum_{\bfQ, \bfQ'}& \int \rmd\bfq\; \mathcal{D}(\bfQ - \bfq) \mathcal{D}(\bfQ' - \bfq)\chi(\bfq)\,.
\end{split}
\end{equation}

From \cref{eq:phase_boundary}, we obtain the critical pump strength $\Omega_\rmd$ and study its statistics over independent disorder realizations.
We find that, in the limits $\xi\to0$ and $\xi\to\infty$, realization-to-realization fluctuations in the phase boundary vanish, and we obtain a deterministic expression for the critical drive strength
\begin{equation}\label{eq:phase_boundary_disorder_avg}
    {\Omega_\rmd(\Delta_{\rmc\rmd})} = \frac{\Delta_{\rma\rmd}}{\Omega\mu}\sqrt{\frac{-32}{V\sum_{\bfQ}\chi(\bfQ)}\left(\Delta_{\rmc\rmd}+\frac{\kappa^2}{4\Delta_{\rmc\rmd}}\right)},
\end{equation}
where $V$ is the system volume, and
\begin{equation}\label{eq:mu}
\mu = \overline{\frac{1}{1 + I(\bfr)/\langle I \rangle}}.
\end{equation}
In the previous equation and in the rest of the manuscript we denote with $\overline{\;\cdot\;}$ the disorder average.
\cref{eq:phase_boundary_disorder_avg} describes the clean-system phase boundary with an effective light-matter coupling renormalized by a factor $\mu$.
As shown in Appendix~\ref{app:appendix_A}, this factor depends on the speckle correlation length and ranges from $\mu =\int_0^\infty e^{-x}/(1+x)\,\rmd x\simeq 0.6$ for $\xi \to 0$ to $\mu = 0.5$ for $\xi \to \infty$.
 While vanishing fluctuations in the $\xi\to\infty$ limit are expected (when the disorder correlation length is much larger than the system size, the system is effectively clean), zero disorder variance when $\xi\to0$ reveals the existence of self-averaging behavior. 
Since $\mu>0.5$ in this limit, we conclude that self-averaging disorder lowers the DWO threshold relative to the clean case [that is, when $I(\bfr)$ is constant]. 
Numerical realizations of the speckle disorder further show that this behavior persists for all finite $\xi$.
This enhancement arises because the disordered coupling couples the cavity field to a continuum of density Fourier components rather than to the discrete set selected by the clean cavity geometry.
The harmonic trap is incorporated through a local density approximation, as detailed in Appendix~\ref{app:appendix_A}.

To complement the analytical linear response analysis, we study the DWO transition with a numerical mean-field treatment.
This approach gives access to quantities beyond the linear response threshold, in particular the coherent cavity amplitude $\alpha = \expval{\hat a}$ and the real-space fermionic density $n(\bfr) =\expval{\hat n(\bfr)}$.
The time-domain mean-field equations of motion are
\begin{equation}\label{eqs:mf_eom}
\begin{split}
    &\frac{\partial\alpha}{\partial t} = -\left(\rmi\Delta_{\rmc\rmd}+\frac{\kappa}{2}\right)\alpha - \rmi\Tr\left(g^\top \rho\right),\\
    &\frac{\partial \rho_{jk}}{\partial t} = \rmi[h_{\rm eff},\,\rho]_{jk}.
\end{split}
\end{equation}
Here $\rho_{jk} = \expval*{\hat c_j^\dagger \hat c_k}$ are the single-particle density matrix elements in the two-dimensional harmonic-oscillator basis, and $(h_{\rm eff})_{jk} = \delta_{jk} \varepsilon_j + 2\operatorname{Re}(\alpha) g_{kj}$, with $\varepsilon_j$ the trap eigenenergies.
The disordered light-matter coupling is encoded in the matrix
\begin{equation}
    g_{jk}=\frac{\Omega\Omega_\rmd}{2\Delta_{\rma\rmd}}\int \rmd\bfr\, \frac{\cos(\bfk_\rmpp\cdot\bfr)\cos(\bfk_\rmc\cdot\bfr)\varphi_j^*(\bfr)\varphi_k(\bfr)}{1+I(\bfr)/\langle I \rangle},
\end{equation}
where $\varphi_j(\bfr)$ are harmonic-oscillator eigenfunctions.
Steady states are found by imposing $\partial\alpha/\partial t=\partial \rho_{jk}/\partial t=0$ and solving the resulting self-consistency equations for $\operatorname{Re}(\alpha)$ and $\rho_{jk}$.

\section{Density wave ordering with controllable disorder}\label{sec:results}

DWO transitions appear simultaneously in the photonic degree of freedom, through the buildup of a macroscopic coherent cavity population, and in the fermionic degree of freedom, through the organization of the atoms into a crystalline density pattern~\cite{mivehvar_cavity_2021}.
Both signatures are experimentally accessible: the former through photons leaking out of the cavity~\cite{helson_optomechanical_2022, helson_density-wave_2023, wu_signatures_2023, zhang_observation_2021, zwettler_nonequilibrium_2025}, and the latter through \emph{in situ} imaging of the fermionic cloud~\cite{buhler_microscopy_2026}.
In this section, we study how speckle disorder modifies the DWO transition and the self-organized phase from both perspectives.

\subsection{Photonic observables}\label{sec:photons}

We first study the mean steady-state intracavity photon number $|\alpha|^2$ as a function of the drive strength $\Omega_\rmd$ and cavity-drive detuning $\Delta_{\rmc\rmd}$.
We choose a speckle realization with correlation length $\xi=0.1\mu\rm m$, for which the system lies in the self-averaging regime and thus single-realization results are representative.
\cref{fig:phase_diagrams clean,fig:phase_diagrams disordered} compare the phase diagrams of a disordered system and of the corresponding clean one obtained with a constant $I(\bfr)$, for $N=196$ fermionic modes at half filling.
In both cases, the analytical phase boundary from \cref{eq:phase_boundary} agrees well with the abrupt onset of a finite photon number obtained from the mean-field steady states.
The comparison also shows that disorder favors the superradiant transition: at fixed detuning, the critical drive strength is lower in the disordered system than in the clean one (cf. \cref{fig:phase_diagrams photons}).

Next, we analyze the dependence of the superradiant threshold on the speckle correlation length.
For each value of $\xi$, we generate $1000$ disorder realizations and compute the critical $\Omega_\rmd$ from  \cref{eq:phase_boundary}.
The disorder-averaged threshold and its standard deviation are shown in \cref{fig:disorder_avg_var threshold,fig:disorder_avg_var variance}.
In the limit $\xi\to0$, the fermionic cloud samples many independent speckle grains, yielding a self-averaging threshold with vanishing realization-to-realization fluctuations. 
The corresponding value of $\mu$ computed according to \cref{eq:mu} approaches the analytical prediction of $\mu \simeq 0.6$. 
In the opposite limit $\xi\to\infty$, the speckle profile becomes effectively constant across the cloud and the clean threshold is recovered ($\mu \to 0.5$).
This behavior is due to the renormalization of the speckle $I(\bfr)$ by its mean intensity $\langle I(\bfr) \rangle$ in \cref{eq:int_hamiltonian}.
As we specified in \cref{sec:model}, this speckle renormalization choice allows us to isolate the contribution coming from the spatial randomness only, rather than from the mean, random shift of the atom-to-drive detuning $\Delta_{\rmd\rma}$.

The strongest sample-to-sample fluctuations induced by spatial randomness, quantified in \cref{fig:disorder_avg_var variance} by the variance of the critical drive amplitude $\Omega_\rmd$ over the disorder realizations, occurs at intermediate correlation lengths, where the spatial structure of the speckle pattern is resolved by the atomic cloud.
Furthermore, in \cref{fig:disorder_avg_var} we also study the dependence of $\overline{\Omega}_\rmd$ and its variance with the number of fermionic modes $N$.
While the behavior of $\overline{\Omega}_\rmd$ appears robust with $N$, $\textrm{var}[\Omega_\rmd]$ is suppressed by the particle number.
Large fluctuations are therefore more visible in the mesoscopic regime of few hundreds of fermionic modes.

\subsection{Fermionic observables}\label{sec:fermions}

The fermionic density $n(\bfr)$ is obtained from the steady-state single-particle density matrix $\rho_{jk}$ as
\begin{equation}
    n(\bfr)=\sum_{jk}\varphi_j^*(\bfr)\varphi_k(\bfr)\rho_{jk}.
\end{equation}
\cref{fig:DWO_transition} shows the evolution of this density across the transition.
Below threshold, $n(\bfr)$ has no DWO modulation.
Above threshold, a clean system develops the characteristic periodic DWO lattice, with spatial oscillations set by the wave vectors $\bfQ$.
For the orthogonal pump-cavity geometry considered here, this produces a regular checkerboard pattern.
In the disordered system, the transition still occurs, but the DWO lattice is visibly distorted by the spatially random AC-Stark shift.
This follows directly from the momentum structure of the coupling: in the clean system, only density components at $\bfQ$ couple to the cavity and acquire a finite amplitude, whereas disorder also couples density modes with $\bfq\neq\bfQ$.
These additional Fourier components generate spatial modulations at wavelengths different from those of the clean optical lattice, producing a distorted DWO state.

The magnitude of this distortion depends strongly on the speckle correlation length. For small $\xi$, the system is self-averaging and the speckle has little visible effect on the DWO crystal. For $\xi\to\infty$, the coupling becomes effectively uniform and the clean lattice is recovered. The strongest distortions therefore occur at intermediate correlation lengths, comparable to the wavelengths associated with the DWO state of a clean system. To see this behavior, we compute the density structure factor
\begin{equation}
S(\bfq)=\left|\int n(\bfr)\rme^{\rmi\bfq\cdot\bfr}\,\rmd\bfr\right|^2.
\end{equation}
\cref{fig:structure_factor clean,fig:structure_factor disordered} compare the structure factor of a clean system $S_\rmc(\bfq)$ and that of a disordered system $S_\rmd(\bfq)$. The former is symmetric and regular, with well-defined peaks at the origin and at the ordering wavevectors $\bfQ$. In contrast, $S_\rmd(\bfq)$ displays disorder-induced distortions, which become apparent when considering the difference between the two. The most prominent effect is a suppression of the peaks at $\bfQ$, as spectral weight is redistributed to other wavevectors in a disordered manner, governed by the $\mathcal{D}(\bfq)$ function. 

We quantify the deviation from the clean DWO pattern using the normalized $L^2$ distance
\begin{equation}
\|S_\mathrm{d}(\bfq)-S_\mathrm{c}(\bfq)\|_{L^2}=\left(\int|S_\mathrm{d}(\bfq)-S_\mathrm{c}(\bfq)|^2\,\rmd\bfq\right)^{1/2},
\end{equation}
rescaled by the Fourier-space area, i.e. we calculate the root mean squared difference (RMSD) between the two structure factors.
As visible in \cref{fig:structure_factor rmsd}, this distance is non-monotonic in $\xi$: it is small in both the $\xi\to0$ and $\xi\to\infty$ limits and peaks around $\xi\sim \lambda_p$.
Again, the latter behavior is due to the speckle intensity renormalization, as discussed in \cref{sec:model,sec:photons}.
The distortion also decreases with the intracavity photon number, that is, deeper in the superradiant phase.

\section{Conclusion and outlook}\label{sec:conclusion}

In this work, we studied how a controllable disordered light-matter coupling modifies density-wave ordering in a fermionic cavity-QED system.
The disorder considered here is not introduced as a static potential acting directly on the atoms, but rather as a spatial modulation of the atom-light coupling generated by a speckle-induced AC-Stark shift.
This distinction is important: in momentum space, the clean cavity geometry selects only the density modes at $\bfQ=\pm(\bfk_\rmc\pm\bfk_\rmpp)$, whereas the speckle modulation convolves this coupling with the disorder spectrum and thereby connects the cavity field to a continuum of fermionic density fluctuations.
As a result, engineered disorder can participate directly in the collective feedback mechanism responsible for superradiance.

Our linear response analysis shows that this additional momentum mixing changes the superradiant threshold in a systematic way.
For the parameters considered, the disorder lowers the critical pump strength relative to the corresponding clean system, which can be understood as an effective enhancement of the fermion-cavity response through the coupling to additional density modes.
The dependence on the speckle correlation length reveals two limiting regimes.
When the disorder varies on length scales much shorter than the fermion cloud, the system samples many independent speckle grains and the transition becomes self-averaging, so that different disorder realizations give the same threshold in the thermodynamic limit.
When the speckle varies only on scales much larger than the cloud, the coupling becomes effectively spatially uniform and the clean behavior is recovered.
Between these limits, sample-to-sample fluctuations provide an additional experimentally accessible signature of the finite spatial structure of the disorder.

The mean-field calculations complement this threshold analysis by showing how the ordered state itself is modified above the transition.
The photonic response remains a global and sensitive probe of the onset of order, through the appearance of a finite coherent cavity amplitude and a macroscopic photon number.
At the same time, the fermionic density reveals the microscopic consequence of the disordered coupling: the clean checkerboard-like DWO crystal is distorted by the occupation of Fourier components with $\bfq\neq\bfQ$.
This distortion vanishes both in the self-averaging limit and in the effectively clean limit, and it is strongest at intermediate correlation lengths where the speckle grains are resolved over the spatial extent of the cloud.
The structure factor of the density therefore provides a direct way of quantifying how engineered disorder reshapes the self-organized fermionic crystal.

These results suggest that cavity-coupled fermions can serve not only as a platform for creating ordered states, but also as a diagnostic tool for disordered quantum matter.
The light leaking from the cavity gives access to disorder-induced shifts and fluctuations of the collective instability~\cite{helson_density-wave_2023}, while \emph{in situ} imaging~\cite{buhler_microscopy_2026} probes the spatial reorganization of the atomic density after the transition.
Because the correlation length of the speckle pattern can be tuned independently of the pump strength and cavity detuning, the setup offers a route to separate the effects of disorder on the critical point from its effects on the structure of the ordered phase.
This separation should be useful for future experiments aiming to characterize disordered long-range interacting fermions through both optical and density-resolved observables in the mesoscopic regime of few hundreds of atoms~\cite{orsi2026fermipressureassistedcavitysuperradiancemesoscopic}.

Several extensions follow naturally from this work.
The present analysis focused on non-interacting fermions in a harmonic trap, treating the ordered phase at the mean-field level.
Including fermion-fermion interactions, such as those giving rise to the BEC-BCS crossover~\cite{giorgini_theory_2008, zwerger_bcs-bec_2012}, is an important next step, as interactions can modify the density response that enters the instability, compete or cooperate with the cavity-induced ordering, and generate additional correlated phases in the presence of disorder.
It would also be interesting to study finite-temperature effects, the real-time dynamics of the transition, and regimes where fluctuations beyond mean field become important.
More broadly, combining controllable disorder with cavity-mediated interactions opens a path toward exploring how localization, thermalization, and collective symmetry breaking coexist in driven-dissipative fermionic systems.

\begin{acknowledgments}
We acknowledge useful discussions with Francesca Orsi, Gaia Stella Bolognini and Jean-Philippe Brantut.
We acknowledge support from the Swiss National Science Foundation through Projects No. 200020\_215172, 200021-227992, and 20QU-1\_215928, and as a part of NCCR SPIN (grant number 225153).
\end{acknowledgments}

\appendix

\begin{widetext}

\section{Details on methods}\label{app:appendix_A}

\subsection{Derivation of the DWO threshold}

In this section we detail the derivation of the phase-boundary condition in \cref{eq:phase_boundary} from the cavity-QED Hamiltonian in \cref{eq:cQED_Hamiltonian}.
First, we write the interaction Hamiltonian in \cref{eq:int_hamiltonian} in terms of single-particle momentum eigenstates, with annihilation (creation) operators $\hat{c}_\bfk$ ($\hat{c}^\dagger_\bfk$).
Using $\hat{\Psi}(\bfr)=\int \frac{\rmd\bfk}{2\pi}\,\rme^{\rmi\bfk\cdot \bfr} \hat{c}_{\bfk}$
we obtain
\begin{equation}
    H_\mathrm{int}=\frac{\Omega_\rmd\Omega}{4\sqrt{2}\Delta_{\rma\rmd}}\hat{x}\sum_{\bfQ} \iiint  \frac{\rmd\bfk \, \rmd\bfk'}{(2\pi)^2}\rmd\bfr\,\frac{\rme^{\rmi\bfQ\cdot\bfr} \, \rme^{-\rmi\bfk\cdot\bfr} \, \rme^{\rmi\bfk'\cdot\bfr}}{1+I(\bfr)/\langle I\rangle}  \; \hat{c}^\dagger_{\bfk} \, \hat{c}_{\bfk'} \, .
\end{equation}
Here we have expanded $\cos(\bfk\cdot\bfr)=(\rme^{\rmi\bfk\cdot\bfr}+\rme^{-\rmi\bfk\cdot\bfr})/2$ and introduced the four wave vectors $\bfQ=\pm\bfk_\rmc\pm\bfk_\rmpp$.
We recognize the integral over $\bfr$ as the Fourier transform of the speckle-dependent coupling evaluated at $\bfQ+\bfk'-\bfk$.
Defining $\mathcal{D}(\bfq)=\mathcal{F}\qty[1/\qty(1+I/\langle I\rangle)]$ and changing variables to $\bfk-\bfk'=\bfq$ and $\bfk'=\bfp$, we find
\begin{align}
    H_\mathrm{int}= \eta\,\hat{x}\sum_{\bfQ} \iint\frac{\rmd\bfp}{2\pi} \, \rmd\bfq\, \mathcal{D}(\bfQ-\bfq) \; \hat{c}^\dagger_{\bfp+\bfq} \, \hat{c}_{\bfp}
    = \eta\,\hat{x}\sum_{\bfQ} \int\rmd\bfq\, \mathcal{D}(\bfQ-\bfq) \; \hat{\rho}_{\bfq},
\end{align}
where $\eta=\Omega_\rmd\Omega/(\sqrt{32}\Delta_{\rma\rmd})$ and $\hat{\rho}_{\bfq}=\frac{1}{2\pi}\int\rmd\bfp\,\hat{c}^\dagger_{\bfp+\bfq} \, \hat{c}_{\bfp} $ is the Fourier transform of the density operator.
Together with the Lindblad jump operator $\hat{L}=\sqrt{\kappa}\hat{a}$, this Hamiltonian gives the cavity equations of motion in \cref{eqs:cavity_eom}.
The fermionic equations of motion are not closed at the operator level.
In particular, the equation for $\hat{c}^\dagger_{\bf{p}+\bf{q}} \hat{c}_{\bf{p}}$ reads
\begin{align}
    \frac{\partial}{\partial t} \hat{c}^\dagger_{\bfp+\bfq} \hat{c}_{\bfp}
    = (\varepsilon_{\bfp+\bfq} - \varepsilon_{\bfp}) \hat{c}^\dagger_{\bfp+\bfq} \hat{c}_{\bfp} + \frac{\eta}{2\pi}\hat{x}\sum_{\bfQ} \int \rmd\bfq' \, \mathcal{D}(\bfQ - \bfq')
    \left( \hat{c}^\dagger_{\bfp+\bfq+\bfq'} \hat{c}_{\bfp} - \hat{c}^\dagger_{\bfp+\bfq} \hat{c}_{\bfp-\bfq'} \right).
\end{align}
Since our ultimate goal is to analyze the instabilities of this dynamical system, we linearize this equation around the fermionic state below the phase transition -- a translationally invariant Fermi sea.
Taking expectation values over this state and transforming to the frequency domain yields
\begin{equation}
    \omega \langle\hat{c}^\dagger_{\bfp+\bfq} \hat{c}_{\bfp}\rangle = (\varepsilon_{\bfp+\bfq} - \varepsilon_{\bfp}) \langle\hat{c}^\dagger_{\bfp+\bfq} \hat{c}_{\bfp}\rangle - \frac{\eta}{2\pi} x\sum_{\bfQ} \mathcal{D}(\bfQ - \bfq) (n_{\bfp} - n_{\bfp+\bfq})\,,
\end{equation}
where $n_{\bfk}=\langle\hat{c}^\dagger_{\bfk}\hat{c}_{\bfk}\rangle$.
After integration over $\bfp$, this becomes
\begin{align}
    \rho_{\bfq} = \eta x \sum_{\bfQ} \mathcal{D}(\bfQ - \bfq) \frac{1}{(2\pi)^2}\int \rmd\bfp \, \frac{n_{\bfp+\bfq} - n_{\bfp}}{\varepsilon_{\bfp+\bfq} - \varepsilon_{\bfp} - \omega}= \eta\sum_{\bfQ} \mathcal{D}(\bfQ - \bfq) \, \chi(\bfq, \omega)x\,,
\end{align}
which identifies $\chi(\bfq,\omega)$ as the Lindhard function, the density-density response function of the noninteracting Fermi gas~\cite{giuliani_quantum_2005}.
The linearized dynamics is therefore governed by the system of equations
\begin{equation}
\begin{cases}
    \rmi\omega x=\Delta_{\rmc\rmd}{p}-\dfrac\kappa2 x, \\
    \rmi\omega p=-\Delta_{\rmc\rmd}x-\dfrac\kappa2 p - \eta\displaystyle\sum_{\bfQ} \int \mathcal{D}(\bfQ-\bfq) \, {\rho}_{\bfq} \, \rmd\bfq, \\
    \rho_{\bfq} =  \eta\displaystyle\sum_{\bfQ} \mathcal{D}(\bfQ - \bfq) \, \chi(\bfq, \omega)x.
\end{cases}
\end{equation}
A nontrivial solution of this system exists only if
\begin{equation}
\label{eq:frequency_eq}
\left(\frac\kappa2 + \rmi\omega\right)^2
=-\Delta_{\rmc\rmd}^2 - \eta^2 \Delta_{\rmc\rmd} \sum_{\bfQ, \bfQ'} \int \rmd\bfq\,\chi(\bfq, \omega) \, \mathcal{D}(\bfQ - \bfq) \, \mathcal{D}(\bfQ' - \bfq)
\end{equation}
is satisfied.
This equation determines the frequency of the collective photon-fermion mode within linear response theory.

Following the procedure of Refs.~\cite{zwettler_nonequilibrium_2025, marijanovic_quench_2026}, generalized here to a disordered coupling, we search for purely imaginary solutions $\omega=\rmi\Gamma$ of \cref{eq:frequency_eq}.
These solutions describe exponential growth or decay of the cavity quadratures and hence of the intracavity photon number.
The growth rate $\Gamma$ satisfies
\begin{equation}
    \begin{aligned}
        1 + \frac{(\Gamma-\kappa/2)^2}{\Delta_{\rmc\rmd}^2}
        = -\frac{\eta^2}{\Delta_{\rmc\rmd}} \sum_{\bfQ, \bfQ'} \int \rmd\bfq\; \mathcal{D}(\bfQ - \bfq) \mathcal{D}(\bfQ' - \bfq)
        \int_{-\infty}^{+\infty} \frac{\rmd\omega}{\pi} \, \frac{\omega \, \operatorname{Im} \chi(\bfq, \omega)}{\omega^2 + \Gamma^2}.
    \end{aligned}
\end{equation}
The phase transition occurs when the growth rate crosses zero: below threshold, fluctuations in the fermion density and cavity field decay, whereas above threshold they grow exponentially and drive the system into the ordered superradiant state.
Setting $\Gamma=0$ and using the Kramers-Kronig relation for the frequency integral yields
\begin{align}
    1 + \frac{(\kappa/2)^2}{\Delta_{\rmc\rmd}^2}
    = -\frac{\eta^2}{\Delta_{\rmc\rmd}} \sum_{\bfQ, \bfQ'} \int \rmd\bfq\; \mathcal{D}(\bfQ - \bfq) \mathcal{D}(\bfQ' - \bfq)\chi(\bfq,0)\,.
\end{align}
This equation is equivalent to \cref{eq:phase_boundary} and defines the boundary between the normal and superradiant/DWO phases.

\subsection{Disorder average and variance of the phase boundary}

We now study the statistics of the threshold over disorder realizations, with the goal of deriving \cref{eq:phase_boundary_disorder_avg} and characterizing how the mean and variance of the critical point depend on the speckle correlation length.
We first decompose the disorder contribution as
\begin{align}
    \mathcal{D}(\bfk)&=\mathcal{F}\left[\mu+\frac{1}{1+I(\bfr)/\langle I(\bfr)\rangle}- \mu\right](\bfk) =2\pi\mu\delta(\bfk)+\mathcal{F}\left[\frac{1}{1+I(\bfr)/\langle I(\bfr)\rangle}- \mu\right](\bfk)\equiv2\pi\mu\delta(\bfk)+\tilde{\mathcal{D}}(\bfk),
\end{align}
where
\begin{equation}
    \mu\equiv\overline{\frac{1}{1+I(\bfr)/\langle I(\bfr)\rangle}}
\end{equation}
and the overline denotes the disorder average.
Substitution into \cref{eq:phase_boundary} gives
\begin{align}
    \Delta_{\rmc\rmd}+\frac{(\kappa/2)^2}{\Delta_{\rmc\rmd}}=
    \eta^2\bigg[V\mu^2\sum_{\bfQ}\chi(\bfQ)
    +4\pi^2\mu\sum_{\bfQ,\bfQ'}\tilde{\mathcal{D}}(\bfQ-\bfQ')\chi(\bfQ)+\sum_{\bfQ, \bfQ'} \int \rmd\bfq\; \tilde{\mathcal{D}}(\bfQ - \bfq) \tilde{\mathcal{D}}(\bfQ' - \bfq)\chi(\bfq)\bigg].
\end{align}
The first term is the phase boundary of a clean system with the coupling rescaled by $\mu^2$.
For a given disorder realization, define
\begin{align}
    D[I(\bfr)]= \frac{1}{V\mu^2\sum_{\bfQ}\chi(\bfQ)}
    \bigg[4\pi^2\mu\sum_{\bfQ,\bfQ'}\tilde{\mathcal{D}}(\bfQ-\bfQ')\chi(\bfQ)+\sum_{\bfQ, \bfQ'} \int \rmd\bfq\; \tilde{\mathcal{D}}(\bfQ - \bfq) \tilde{\mathcal{D}}(\bfQ' - \bfq)\chi(\bfq)\bigg].
\end{align}
The critical drive strength can then be written as
\begin{equation}
    {\Omega_\rmd}(\Delta_{\rmc\rmd}) = \frac{\Delta_{\rma\rmd}}{\Omega\mu}\sqrt{\frac{32}{ V\sum_{\bfQ}  [-\chi(\bfQ)]}\left(\Delta_{\rmc\rmd}+\frac{\kappa^2}{4\Delta_{\rmc\rmd}}\right)\frac{1}{1+D}}\,.
\end{equation}
Expanding $(1+D)^{-1/2}$ gives the disorder-averaged threshold in terms of the moments of the random variable $D$,
\begin{equation}
    \overline{\Omega_\rmd}(\Delta_{\rmc\rmd}) = \frac{\Omega_\rmd^\mathrm{(clean)}(\Delta_{\rmc\rmd})}{\mu}\left(1-\frac{1}{2}\overline{D}+\frac{3}{8}\overline{D^2}+\mathcal O(\overline{D^3})\right),
\end{equation}
where $\Omega_\rmd^\mathrm{(clean)}(\Delta_{\rmc\rmd})$ is the critical drive strength of the clean system.
An exact expression would require all moments of $D$; instead, we focus on the two limiting regimes.
For $\xi\to\infty$, the coupling is effectively uniform, $I(\bfr)=\langle I(\bfr)\rangle$, so $D$ and all of its moments vanish.
More generally, the first moment reads
\begin{align}
    \overline{D}= \frac{1}{V\mu^2\sum_{\bfQ}\chi(\bfQ)}
    \bigg[&4\pi^2\mu\sum_{\bfQ,\bfQ'}\overline{\tilde{\mathcal{D}}(\bfQ-\bfQ')}\chi(\bfQ)+\sum_{\bfQ, \bfQ'} \int \rmd\bfq\; \overline{\tilde{\mathcal{D}}(\bfQ - \bfq) \tilde{\mathcal{D}}(\bfQ' - \bfq)}\chi(\bfq)\bigg].
\end{align}
The first term vanishes because $\overline{\tilde{\mathcal{D}}(\bfk)}=0$.
The second term is determined by Fourier transforms of correlators of the disorder at different points,
\begin{align}
    \overline{\tilde{\mathcal{D}}(\bfQ - \bfq) \tilde{\mathcal{D}}(\bfQ' - \bfq)}
    \quad=\iint\rmd\bfr\rmd\bfr'\,\overline{\left(\frac{1}{1+I(\bfr)/\langle I(\bfr)\rangle}-\mu \right)
    \left(\frac{1}{1+I(\bfr\,')/\langle I(\bfr\,')\rangle}-\mu \right)} e^{\rmi(\bfQ-\bfq)\cdot\bfr+\rmi(\bfQ'-\bfq)\cdot\bfr'}.
\end{align}
These correlators vanish in the $\xi\to0$ limit as distinct spatial points become uncorrelated.
The same argument applies to higher moments of $D$, which involve higher-order disorder correlators.
Thus, in both limiting regimes considered here,
\begin{equation}
    \overline{\Omega_\rmd}(\Delta_{\rmc\rmd})\big|_{\xi\to0,\infty} = \frac{\Omega_\rmd^\mathrm{(clean)}(\Delta_{\rmc\rmd})}{\mu\big|_{\xi\to0,\infty}}\,.
\end{equation}

The variance follows from a similar expansion,
\begin{align}
    \mathrm{Var}[{\Omega_\rmd}(\Delta_{\rmc\rmd})]
    = \frac{[\Omega_\rmd^\mathrm{(clean)}(\Delta_{\rmc\rmd})]^2}{\mu^2}
    \bigg[&\frac{1}{4}\left(\overline{D^2}-\overline{D}^2\right)
    -\frac38\left(\overline{D^3}-\overline D\,\overline{D^2}\right)
    +\mathcal O(\overline{D^4})\bigg].
\end{align}
As before, this trivially approaches $0$ in the $\xi\to\infty$ limit as $D$ vanishes. In the opposite limit, $\xi\to0$, the variance also vanishes as the average and all higher-order moments of $D$ go to 0, reflecting a self-averaging behavior.

It remains to determine $\mu$ in the two limits.
For $\xi\to\infty$, $I(\bfr)=\langle I(\bfr)\rangle$ and therefore $\mu_{\xi\to\infty}=1/2$.
For $\xi\to0$, we use the fact that speckle intensities follow an exponential distribution, equivalently a Gamma distribution with shape parameter $\alpha=1$: $I(\bfr)\sim\mathrm{Gamma}(1,\lambda)$.
We discretize the speckle field into $N$ independent cells at positions $\bfr_i$, over which the intensity is approximately constant allowing us to treat the disorder as a set of independent identically distributed random variables.
Then
\begin{equation}
    \frac{I(\bfr)}{\langle I(\bfr)\rangle}\approx N\frac{I(\bfr_i)}{\sum_jI(\bfr_j)}\,.
\end{equation}
If $X_i\,,\,i=1,...,N$ are a set of Gamma distributed random variables with common scale parameter $\lambda$, then the random variables $X_i/\sum_jX_j$ follow a Dirichlet distribution, $\mathrm{Dirichlet}(\alpha_1,...,\alpha_N)$, whose marginal is a Beta distribution $\mathrm{Beta}(\alpha_i,\sum_{j\neq i}\alpha_j)$.
Thus
\begin{equation}
    \frac{I(\bfr)}{N\langle I(\bfr)\rangle}=P\sim \mathrm{Beta}(1,N-1),
\end{equation}
and the normalized intensity has probability density
\begin{equation}
    f(p) = \frac{\Gamma(N)}{\Gamma(1)\Gamma(N-1)}(1-p)^{N-2}= (N-1)(1-p)^{N-2}.
\end{equation}
We therefore obtain
\begin{equation}
    \mu = \int_0^1 \rmd p\, \frac{1}{1 + N p} (N-1)(1-p)^{N-2} \,.
\end{equation}
In the $\xi\to0$ limit, $N\to\infty$, and
\begin{align}
    \mu = \lim_{N\to\infty}\int_0^1\rmd p\, \frac{1}{1 + N p} (N-1)(1-p)^{N-2}\overset{s=Np}{=}\lim_{N\to\infty}\int_0^N \rmd s\, \frac{1}{1 + s} \left(1-\frac{s}{N}\right)^{N-2} =\int_0^\infty\rmd s\,\frac{e^{-s}}{1+s}\simeq0.6.
\end{align}
The last step follows from dominated convergence.
The limiting thresholds are therefore
\begin{equation}
\begin{aligned}
\label{eq:sel_avg_thresh}
    {\Omega_\rmd}(\Delta_{\rmc\rmd})\bigg|_{\xi\to0}
    = \frac{\Delta_{\rma\rmd}}{\Omega\mu_0}\sqrt{\frac{32}{ V\sum_{\bfQ}[-\chi(\bfQ)]}\left(\Delta_{\rmc\rmd}+\frac{\kappa^2}{4\Delta_{\rmc\rmd}}\right)},\qquad\qquad
    \mu_0=\int_0^\infty\rmd s\,\frac{e^{-s}}{1+s}\simeq0.6,
\end{aligned}
\end{equation}
and
\begin{equation}
\label{eq:clean_thresh}
    {\Omega_\rmd}(\Delta_{\rmc\rmd})\bigg|_{\xi\to\infty}
    = \frac{2\Delta_{\rma\rmd}}{\Omega}\sqrt{\frac{32}{ V\sum_{\bfQ}[-\chi(\bfQ)]}\left(\Delta_{\rmc\rmd}+\frac{\kappa^2}{4\Delta_{\rmc\rmd}}\right)}.
\end{equation}
The latter is the threshold of a clean system with spatially constant detuning beam, resulting in a doubling of the atom-drive detuning.
Since $1/\mu_0<2$, the self-averaged short-correlation-length disorder lowers the pump strength required to reach the transition relative to this clean limit.
In the crossover between the two limits, an analytical treatment is less direct; the numerical results in the main text indicate that the disorder-averaged threshold varies monotonically with $\xi$, therefore always remaining on average below the clean system's, while the variance peaks at intermediate correlation length.

\end{widetext}

\subsection{Trap averaging}

\begin{figure*}[t]
    \centering
    \includegraphics[width=\linewidth]{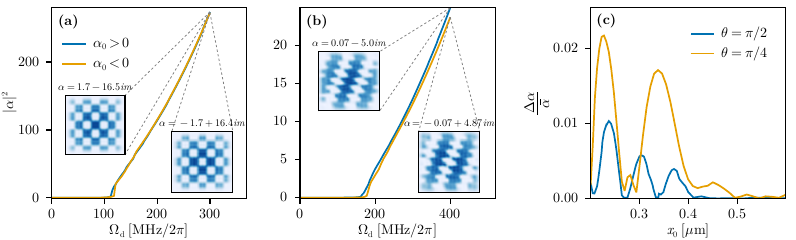}
    \caption{(a-b) Cavity photon number as a function of $\Omega_\rmd$ for the two mean-field fixed points related by the $\mathbb{Z}_2$ symmetry. The two branches are obtained from initial conditions with positive and negative real values of the cavity field $\alpha_0$. Insets: fermion density above the threshold for the two solutions and the corresponding coherent cavity amplitude $\alpha$. Results are shown for two angles between the drive beam and the cavity axis: (a) $\theta = \pi/2$ and (b) $\theta = \pi/4$. (c) Difference between the two solutions, normalized by their average, as a function of the harmonic-trap length scale $x_0$ for the two different angles.}
    \label{fig:Z2_symmetry}
    \phantomsubcaptionlabel{fig:Z2_symmetry pi/2}
    \phantomsubcaptionlabel{fig:Z2_symmetry pi/4}
    \phantomsubcaptionlabel{fig:Z2_symmetry diff}
\end{figure*}

To determine the critical threshold, we need the static ($\omega=0$) density-density response function.
For free spin-$1/2$ fermions in two dimensions, the static Lindhard function is
\begin{equation}
\label{eq:Lindhard}
    \chi_\mathrm{free}(q)=-\frac{m_{\rm at}}{\pi\hbar^2}\left[1-\Theta(q-2k_F)\sqrt{1-\left(\frac{2k_F}{q}\right)^2}\right],
\end{equation}
where $m_{\rm at}$ is the atomic mass (throughout the article, we assume $^6$Li atoms) and $k_F$ the Fermi wave vector.
We include the harmonic trap using a local density approximation (LDA).
For a harmonic potential $U(\bfr)=\frac{1}{2}m_{\rm at}\omega_{\rm t}^2r^2$ (being $\omega_{\rm t}=\hbar/m_{\rm t}x_0^2$ the trapping frequency, and $x_0$ the harmonic-trap length), the gas is treated locally as a homogeneous Fermi sea with position-dependent Fermi wave number
\begin{equation}
k_F(\bfr)=\frac{\sqrt{2m_{\rm at}[E_F-U(\bfr)]}}{\hbar}
\end{equation}
inside the region where $E_F>U(\bfr)$ and zero outside it. Here the Fermi energy, $E_F$, is the energy of the highest occupied state, which is taken to be constant in space.
The cloud radius is $R=\sqrt{2E_F/(m_{\rm at}\omega_{\rm t}^2)}$.
The Fermi energy is fixed by the atom number,
\begin{align}
    N&=\int_{|\bfr|<R} \rmd\bfr\,n(\bfr)
      =\int_0^R\rmd r\,\frac{2m_{\rm at}\left(E_F-\frac{1}{2}m_{\rm at}\omega_{\rm t}^2r^2\right)}{\hbar^2}r\nonumber\\
    &\Rightarrow E_F=\hbar\omega_{\rm t}\sqrt{N}\,.
\end{align}
Thus
\begin{equation}
k_F(\bfr)=\frac{\sqrt{2m_{\rm at}\left(\hbar\omega_{\rm t}\sqrt{N}-\frac{1}{2}m_{\rm at}\omega_{\rm t}^2r^2\right)}}{\hbar}
\end{equation}
within the cloud.
The trap-averaged susceptibility is then
\begin{equation}
\chi_\mathrm{HO}(q)=\frac{1}{A}\int\chi_\mathrm{free}[q,k_F(\bfr)]\,\rmd\bfr,
\end{equation}
where $A$ is the area over which the LDA average is taken.

\subsection{Numerical construction of the speckle pattern}

A speckle pattern can be represented as a superposition of plane waves with random relative phases. To generate such a pattern numerically we follow the procedure outlined in Ref.~\cite{speckle_sim}.
Considering a $N_x\times N_y$ two-dimensional grid representing a real-space region of size $L_x\times L_y$,
the corresponding Fourier grid contains points $(2\pi n_x/L_x,2\pi n_y/L_y)$, with integers $n_x$ and $n_y$. To each point in Fourier space, we assign complex amplitudes
\begin{equation}
    E(\bfk)=
    \begin{cases}
        \rme^{2\pi \rmi X}, & |\bfk|\leq 2\pi/\xi,\\
        0, & |\bfk|>2\pi/\xi,
    \end{cases}
\end{equation}
where $X$ is uniformly distributed in $[0,1]$.
The real-space field is obtained from $E(\bfr)=\mathcal{F}^{-1}[E(\bfk)]$, the intensity is $I(\bfr)=|E(\bfr)|^2$, and the pattern is normalized by its spatial average to obtain $I(\bfr)/\langle I(\bfr)\rangle$.
The correlation length $\xi$ sets the typical speckle-grain size: larger $\xi$ produces larger regions of similar intensity.
When $\xi$ becomes comparable to the real-space grid size, the cutoff $|\bfk|\leq2\pi/\xi$ can fall below the Fourier-grid resolution.
To avoid this finite-size artifact, we generate the speckle pattern on a larger grid and then extract a subregion of the desired size, a procedure that also mimics typical experimental implementations.

\section{Impact of the $\mathbb{Z}_2$ symmetry}\label{sec:Z2}

The DWO transition is associated with the breaking of a $\mathbb{Z}_2$ symmetry.
In the absence of the trap, the Hamiltonian in \cref{eq:cQED_Hamiltonian} is invariant under the simultaneous transformation
\begin{equation}
    \hat{a}\to -\hat{a},\qquad \hat{\rho}(\bfQ)\to -\hat{\rho}(\bfQ),
\end{equation}
for the density-wave components coupled to the cavity field.
Physically, changing the sign of the cavity amplitude shifts the phase of the cavity standing wave by $\pi$, exchanging the minima and maxima of the optical lattice.
The fermionic density modulation follows this shift, so the ordered phase has two symmetry-related configurations: one with cavity amplitude $\alpha^*$ and a corresponding DWO pattern, and another with amplitude $-\alpha^*$ and the complementary density pattern.
At the transition, the system spontaneously selects one of these configurations.
This structure is also visible in the numerical mean-field solutions: two stable fixed points are obtained depending on the sign of the initial value of $\alpha$.
\cref{fig:Z2_symmetry pi/2,fig:Z2_symmetry pi/4} show the photon number as a function of the drive strength for the two branches, together with the corresponding fermionic density patterns. These results are presented for two angles between the drive beam and the cavity axis, $\theta$: the orthogonal case ($\theta = \pi/2$), that has been analyzed throughout this work, and $\theta = \pi/4$, which produces a DWO state with different spatial periodicity, since the ordering wavevectors are no longer orthogonal and do not have the same magnitude.

\begin{figure}[t]
    \centering
    \includegraphics[width=\linewidth]{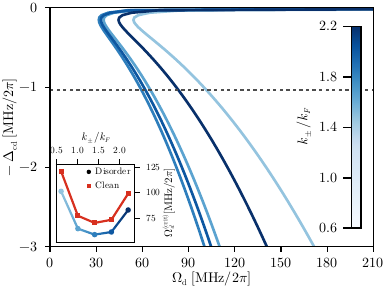}
    \caption{Phase-boundary curves in the $\Omega_\rmd-\Delta_{\rmc\rmd}$ plane, calculated from \cref{eq:phase_boundary}, for different values of the Fermi wave vector $k_F$. The Fermi wavevector is presented in terms of the norm of the DWO wavevectors $k_\pm=|\bfk_\rmc\pm\bfk_\rmpp|$. Inset: Critical value of $\Omega_\rmd$, at fixed $\Delta_{\rmc\rmd}$ (indicated by the dashed line), as a function of the ratio between the norm of the DWO wavevectors and the Fermi wavevector, for a clean and a disordered system with $\xi = 0.1 \mu \mathrm{m}$.
    Parameters as in \cref{fig:phase_diagrams}.}
    \label{fig:kF}
\end{figure}

For free fermions without a trap, the $\mathbb{Z}_2$ symmetry implies that the two solutions have the same $|\alpha|$ and hence the same photon number.
In the trapped system, however, we find a small difference between the photon numbers of the two branches.
This difference originates from the harmonic potential, which weakly breaks the equivalence between the two density-wave configurations: the two complementary patterns place atoms in different regions of the trap and therefore have slightly different potential energies.
As shown in \cref{fig:Z2_symmetry diff}, the difference in $|\alpha|$ generally decreases as the trap becomes wider.
This is expected because, when the trapping potential varies slowly on the scale of the DWO wavelength, the two density configurations become nearly degenerate. 
This decrease is not monotonic, but instead exhibits oscillations whose period can be approximately related to the number of DWO wavelengths that fit inside a radius $R$.
We also find that the difference between the two branches increases for smaller cavity-drive angles, as the potential energy difference between the occupied regions of the two possible ordered states is more evident for DWO crystals characterized by larger wavelengths.

\section{Dependence on the Fermi wave vector}\label{sec:fermi_vector}

\begin{figure}[t]
    \centering
    \includegraphics[width=\linewidth]{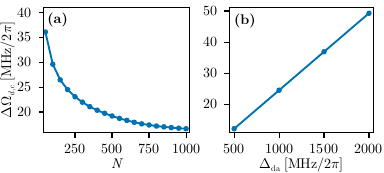}
    \caption{Difference between the superradiant threshold of a clean system and of a disordered system, $\Delta\Omega_{\rm d,c}$ as a function of the (a) number of fermionic modes, $N$ and (b) atom-drive detuning. Other parameters are fixed to $\Delta_{\rmc\rmd}/2\pi = 2\,\textrm{MHz}$, $\Omega/2\pi=4\,\textrm{MHz}$, $\kappa/2\pi=300\,\textrm{kHz}$ and $\xi = 0.1 \mu\mathrm{m}$.}
    \label{fig:DeltaOmega}
    \phantomsubcaptionlabel{fig:DeltaOmega number of fermions}
    \phantomsubcaptionlabel{fig:DeltaOmega detuning}
\end{figure}

In this appendix, we examine how the phase boundary changes with the Fermi wave vector. In the presence of the harmonic trap and within the LDA, $k_F$ becomes spatially dependent. In the following, we take $k_F$ to be its value at the center of the trap $k_F=\sqrt{2m_{\rm at}\omega_{\rm t}\sqrt{N}/\hbar}$. The Fermi wave vector can therefore be tuned by varying the trap frequency or the number of fermionic modes. This modifies the Lindhard response entering \cref{eq:phase_boundary}, and therefore changes the pump strength required to trigger the DWO instability.
\cref{fig:kF} shows the resulting phase-boundary curves in the $\Omega_\rmd-\Delta_{\rmc\rmd}$ plane. We find that the disordered system retains the non-monotonic behavior of the superradiant threshold reported in Ref.~\cite{orsi2026fermipressureassistedcavitysuperradiancemesoscopic}. Initially, the critical point decreases as $k_F$ increases. However, once the excitations that lead to the formation of the DWO crystal become enclosed within the Fermi surface, the critical point begins to increase, indicating a crossover from a regime where Fermi-pressure favors ordering to one in which Pauli exclusion suppresses photon scattering.

\section{Dependence on the atom-drive detuning and the number of fermionic modes}

Finally, we illustrate how the extent to which disorder favors the superradiant phase depends on the system parameters. \cref{fig:DeltaOmega number of fermions} shows the difference between the superradiant threshold of a clean system and that of a disordered system, obtained analytically through \cref{eq:phase_boundary}, as a function of the number of fermionic modes, $N$. As $N$ increases, the drive strength necessary to achieve superradiance decreases for both the clean and disordered systems, thereby reducing the difference between the corresponding thresholds. This reduction becomes progressively slower as $N$ grows. Figure \cref{fig:DeltaOmega detuning}  displays the threshold difference as a function of the atom-drive detuning. In this case, the behavior is straightforward: since the threshold expressions for both the clean and disordered systems depend linearly on $\Delta_{\mathrm{da}}$ [cf. \cref{eq:sel_avg_thresh,eq:clean_thresh}] , their difference also increases linearly with the detuning.


\begin{thebibliography}{54}%
\makeatletter
\providecommand \@ifxundefined [1]{%
 \@ifx{#1\undefined}
}%
\providecommand \@ifnum [1]{%
 \ifnum #1\expandafter \@firstoftwo
 \else \expandafter \@secondoftwo
 \fi
}%
\providecommand \@ifx [1]{%
 \ifx #1\expandafter \@firstoftwo
 \else \expandafter \@secondoftwo
 \fi
}%
\providecommand \natexlab [1]{#1}%
\providecommand \enquote  [1]{``#1''}%
\providecommand \bibnamefont  [1]{#1}%
\providecommand \bibfnamefont [1]{#1}%
\providecommand \citenamefont [1]{#1}%
\providecommand \href@noop [0]{\@secondoftwo}%
\providecommand \href [0]{\begingroup \@sanitize@url \@href}%
\providecommand \@href[1]{\@@startlink{#1}\@@href}%
\providecommand \@@href[1]{\endgroup#1\@@endlink}%
\providecommand \@sanitize@url [0]{\catcode `\\12\catcode `\$12\catcode `\&12\catcode `\#12\catcode `\^12\catcode `\_12\catcode `\%12\relax}%
\providecommand \@@startlink[1]{}%
\providecommand \@@endlink[0]{}%
\providecommand \url  [0]{\begingroup\@sanitize@url \@url }%
\providecommand \@url [1]{\endgroup\@href {#1}{\urlprefix }}%
\providecommand \urlprefix  [0]{URL }%
\providecommand \Eprint [0]{\href }%
\providecommand \doibase [0]{https://doi.org/}%
\providecommand \selectlanguage [0]{\@gobble}%
\providecommand \bibinfo  [0]{\@secondoftwo}%
\providecommand \bibfield  [0]{\@secondoftwo}%
\providecommand \translation [1]{[#1]}%
\providecommand \BibitemOpen [0]{}%
\providecommand \bibitemStop [0]{}%
\providecommand \bibitemNoStop [0]{.\EOS\space}%
\providecommand \EOS [0]{\spacefactor3000\relax}%
\providecommand \BibitemShut  [1]{\csname bibitem#1\endcsname}%
\let\auto@bib@innerbib\@empty
\bibitem [{\citenamefont {Bloch}\ \emph {et~al.}(2008)\citenamefont {Bloch}, \citenamefont {Dalibard},\ and\ \citenamefont {Zwerger}}]{bloch_many-body_2008}%
  \BibitemOpen
  \bibfield  {author} {\bibinfo {author} {\bibfnamefont {I.}~\bibnamefont {Bloch}}, \bibinfo {author} {\bibfnamefont {J.}~\bibnamefont {Dalibard}},\ and\ \bibinfo {author} {\bibfnamefont {W.}~\bibnamefont {Zwerger}},\ }\bibfield  {title} {\bibinfo {title} {Many-body physics with ultracold gases},\ }\href {https://doi.org/10.1103/RevModPhys.80.885} {\bibfield  {journal} {\bibinfo  {journal} {Rev. Mod. Phys.}\ }\textbf {\bibinfo {volume} {80}},\ \bibinfo {pages} {885} (\bibinfo {year} {2008})}\BibitemShut {NoStop}%
\bibitem [{\citenamefont {Gross}\ and\ \citenamefont {Bloch}(2017)}]{gross_quantum_2017}%
  \BibitemOpen
  \bibfield  {author} {\bibinfo {author} {\bibfnamefont {C.}~\bibnamefont {Gross}}\ and\ \bibinfo {author} {\bibfnamefont {I.}~\bibnamefont {Bloch}},\ }\bibfield  {title} {\bibinfo {title} {Quantum simulations with ultracold atoms in optical lattices},\ }\href {https://doi.org/10.1126/science.aal3837} {\bibfield  {journal} {\bibinfo  {journal} {Science}\ }\textbf {\bibinfo {volume} {357}},\ \bibinfo {pages} {995} (\bibinfo {year} {2017})}\BibitemShut {NoStop}%
\bibitem [{\citenamefont {Esslinger}(2010)}]{esslinger_fermi-hubbard_2010}%
  \BibitemOpen
  \bibfield  {author} {\bibinfo {author} {\bibfnamefont {T.}~\bibnamefont {Esslinger}},\ }\bibfield  {title} {\bibinfo {title} {Fermi-{Hubbard} {Physics} with {Atoms} in an {Optical} {Lattice}},\ }\href {https://doi.org/10.1146/annurev-conmatphys-070909-104059} {\bibfield  {journal} {\bibinfo  {journal} {Annual Review of Condensed Matter Physics}\ }\textbf {\bibinfo {volume} {1}},\ \bibinfo {pages} {129} (\bibinfo {year} {2010})}\BibitemShut {NoStop}%
\bibitem [{\citenamefont {Hart}\ \emph {et~al.}(2015)\citenamefont {Hart}, \citenamefont {Duarte}, \citenamefont {Yang}, \citenamefont {Liu}, \citenamefont {Paiva}, \citenamefont {Khatami}, \citenamefont {Scalettar}, \citenamefont {Trivedi}, \citenamefont {Huse},\ and\ \citenamefont {Hulet}}]{hart_observation_2015}%
  \BibitemOpen
  \bibfield  {author} {\bibinfo {author} {\bibfnamefont {R.~A.}\ \bibnamefont {Hart}}, \bibinfo {author} {\bibfnamefont {P.~M.}\ \bibnamefont {Duarte}}, \bibinfo {author} {\bibfnamefont {T.-L.}\ \bibnamefont {Yang}}, \bibinfo {author} {\bibfnamefont {X.}~\bibnamefont {Liu}}, \bibinfo {author} {\bibfnamefont {T.}~\bibnamefont {Paiva}}, \bibinfo {author} {\bibfnamefont {E.}~\bibnamefont {Khatami}}, \bibinfo {author} {\bibfnamefont {R.~T.}\ \bibnamefont {Scalettar}}, \bibinfo {author} {\bibfnamefont {N.}~\bibnamefont {Trivedi}}, \bibinfo {author} {\bibfnamefont {D.~A.}\ \bibnamefont {Huse}},\ and\ \bibinfo {author} {\bibfnamefont {R.~G.}\ \bibnamefont {Hulet}},\ }\bibfield  {title} {\bibinfo {title} {Observation of antiferromagnetic correlations in the {Hubbard} model with ultracold atoms},\ }\href {https://doi.org/10.1038/nature14223} {\bibfield  {journal} {\bibinfo  {journal} {Nature}\ }\textbf {\bibinfo {volume} {519}},\ \bibinfo {pages} {211} (\bibinfo {year} {2015})}\BibitemShut {NoStop}%
\bibitem [{\citenamefont {Mazurenko}\ \emph {et~al.}(2017)\citenamefont {Mazurenko}, \citenamefont {Chiu}, \citenamefont {Ji}, \citenamefont {Parsons}, \citenamefont {Kanász-Nagy}, \citenamefont {Schmidt}, \citenamefont {Grusdt}, \citenamefont {Demler}, \citenamefont {Greif},\ and\ \citenamefont {Greiner}}]{mazurenko_cold-atom_2017}%
  \BibitemOpen
  \bibfield  {author} {\bibinfo {author} {\bibfnamefont {A.}~\bibnamefont {Mazurenko}}, \bibinfo {author} {\bibfnamefont {C.~S.}\ \bibnamefont {Chiu}}, \bibinfo {author} {\bibfnamefont {G.}~\bibnamefont {Ji}}, \bibinfo {author} {\bibfnamefont {M.~F.}\ \bibnamefont {Parsons}}, \bibinfo {author} {\bibfnamefont {M.}~\bibnamefont {Kanász-Nagy}}, \bibinfo {author} {\bibfnamefont {R.}~\bibnamefont {Schmidt}}, \bibinfo {author} {\bibfnamefont {F.}~\bibnamefont {Grusdt}}, \bibinfo {author} {\bibfnamefont {E.}~\bibnamefont {Demler}}, \bibinfo {author} {\bibfnamefont {D.}~\bibnamefont {Greif}},\ and\ \bibinfo {author} {\bibfnamefont {M.}~\bibnamefont {Greiner}},\ }\bibfield  {title} {\bibinfo {title} {A cold-atom {Fermi}–{Hubbard} antiferromagnet},\ }\href {https://doi.org/10.1038/nature22362} {\bibfield  {journal} {\bibinfo  {journal} {Nature}\ }\textbf {\bibinfo {volume} {545}},\ \bibinfo {pages} {462} (\bibinfo {year} {2017})}\BibitemShut {NoStop}%
\bibitem [{\citenamefont {Polkovnikov}\ \emph {et~al.}(2011)\citenamefont {Polkovnikov}, \citenamefont {Sengupta}, \citenamefont {Silva},\ and\ \citenamefont {Vengalattore}}]{polkovnikov_colloquium_2011}%
  \BibitemOpen
  \bibfield  {author} {\bibinfo {author} {\bibfnamefont {A.}~\bibnamefont {Polkovnikov}}, \bibinfo {author} {\bibfnamefont {K.}~\bibnamefont {Sengupta}}, \bibinfo {author} {\bibfnamefont {A.}~\bibnamefont {Silva}},\ and\ \bibinfo {author} {\bibfnamefont {M.}~\bibnamefont {Vengalattore}},\ }\bibfield  {title} {\bibinfo {title} {\textit{{Colloquium}} : {Nonequilibrium} dynamics of closed interacting quantum systems},\ }\href {https://doi.org/10.1103/RevModPhys.83.863} {\bibfield  {journal} {\bibinfo  {journal} {Rev. Mod. Phys.}\ }\textbf {\bibinfo {volume} {83}},\ \bibinfo {pages} {863} (\bibinfo {year} {2011})}\BibitemShut {NoStop}%
\bibitem [{\citenamefont {Bloch}\ \emph {et~al.}(2012)\citenamefont {Bloch}, \citenamefont {Dalibard},\ and\ \citenamefont {Nascimbène}}]{bloch_quantum_2012}%
  \BibitemOpen
  \bibfield  {author} {\bibinfo {author} {\bibfnamefont {I.}~\bibnamefont {Bloch}}, \bibinfo {author} {\bibfnamefont {J.}~\bibnamefont {Dalibard}},\ and\ \bibinfo {author} {\bibfnamefont {S.}~\bibnamefont {Nascimbène}},\ }\bibfield  {title} {\bibinfo {title} {Quantum simulations with ultracold quantum gases},\ }\href {https://doi.org/10.1038/nphys2259} {\bibfield  {journal} {\bibinfo  {journal} {Nat. Phys.}\ }\textbf {\bibinfo {volume} {8}},\ \bibinfo {pages} {267} (\bibinfo {year} {2012})}\BibitemShut {NoStop}%
\bibitem [{\citenamefont {Ritsch}\ \emph {et~al.}(2013)\citenamefont {Ritsch}, \citenamefont {Domokos}, \citenamefont {Brennecke},\ and\ \citenamefont {Esslinger}}]{ritsch_cold_2013}%
  \BibitemOpen
  \bibfield  {author} {\bibinfo {author} {\bibfnamefont {H.}~\bibnamefont {Ritsch}}, \bibinfo {author} {\bibfnamefont {P.}~\bibnamefont {Domokos}}, \bibinfo {author} {\bibfnamefont {F.}~\bibnamefont {Brennecke}},\ and\ \bibinfo {author} {\bibfnamefont {T.}~\bibnamefont {Esslinger}},\ }\bibfield  {title} {\bibinfo {title} {Cold atoms in cavity-generated dynamical optical potentials},\ }\href {https://doi.org/10.1103/RevModPhys.85.553} {\bibfield  {journal} {\bibinfo  {journal} {Rev. Mod. Phys.}\ }\textbf {\bibinfo {volume} {85}},\ \bibinfo {pages} {553} (\bibinfo {year} {2013})}\BibitemShut {NoStop}%
\bibitem [{\citenamefont {Mivehvar}\ \emph {et~al.}(2021)\citenamefont {Mivehvar}, \citenamefont {Piazza}, \citenamefont {Donner},\ and\ \citenamefont {Ritsch}}]{mivehvar_cavity_2021}%
  \BibitemOpen
  \bibfield  {author} {\bibinfo {author} {\bibfnamefont {F.}~\bibnamefont {Mivehvar}}, \bibinfo {author} {\bibfnamefont {F.}~\bibnamefont {Piazza}}, \bibinfo {author} {\bibfnamefont {T.}~\bibnamefont {Donner}},\ and\ \bibinfo {author} {\bibfnamefont {H.}~\bibnamefont {Ritsch}},\ }\bibfield  {title} {\bibinfo {title} {Cavity {QED} with quantum gases: new paradigms in many-body physics},\ }\href {https://doi.org/10.1080/00018732.2021.1969727} {\bibfield  {journal} {\bibinfo  {journal} {Adv. Phys.}\ }\textbf {\bibinfo {volume} {70}},\ \bibinfo {pages} {1} (\bibinfo {year} {2021})}\BibitemShut {NoStop}%
\bibitem [{\citenamefont {Domokos}\ and\ \citenamefont {Ritsch}(2002)}]{domokos_collective_2002}%
  \BibitemOpen
  \bibfield  {author} {\bibinfo {author} {\bibfnamefont {P.}~\bibnamefont {Domokos}}\ and\ \bibinfo {author} {\bibfnamefont {H.}~\bibnamefont {Ritsch}},\ }\bibfield  {title} {\bibinfo {title} {Collective {Cooling} and {Self}-{Organization} of {Atoms} in a {Cavity}},\ }\href {https://doi.org/10.1103/PhysRevLett.89.253003} {\bibfield  {journal} {\bibinfo  {journal} {Phys. Rev. Lett.}\ }\textbf {\bibinfo {volume} {89}},\ \bibinfo {pages} {253003} (\bibinfo {year} {2002})}\BibitemShut {NoStop}%
\bibitem [{\citenamefont {Black}\ \emph {et~al.}(2003)\citenamefont {Black}, \citenamefont {Chan},\ and\ \citenamefont {Vuletić}}]{black_observation_2003}%
  \BibitemOpen
  \bibfield  {author} {\bibinfo {author} {\bibfnamefont {A.~T.}\ \bibnamefont {Black}}, \bibinfo {author} {\bibfnamefont {H.~W.}\ \bibnamefont {Chan}},\ and\ \bibinfo {author} {\bibfnamefont {V.}~\bibnamefont {Vuletić}},\ }\bibfield  {title} {\bibinfo {title} {Observation of {Collective} {Friction} {Forces} due to {Spatial} {Self}-{Organization} of {Atoms}: {From} {Rayleigh} to {Bragg} {Scattering}},\ }\href {https://doi.org/10.1103/PhysRevLett.91.203001} {\bibfield  {journal} {\bibinfo  {journal} {Phys. Rev. Lett.}\ }\textbf {\bibinfo {volume} {91}},\ \bibinfo {pages} {203001} (\bibinfo {year} {2003})}\BibitemShut {NoStop}%
\bibitem [{\citenamefont {Baumann}\ \emph {et~al.}(2010)\citenamefont {Baumann}, \citenamefont {Guerlin}, \citenamefont {Brennecke},\ and\ \citenamefont {Esslinger}}]{baumann_dicke_2010}%
  \BibitemOpen
  \bibfield  {author} {\bibinfo {author} {\bibfnamefont {K.}~\bibnamefont {Baumann}}, \bibinfo {author} {\bibfnamefont {C.}~\bibnamefont {Guerlin}}, \bibinfo {author} {\bibfnamefont {F.}~\bibnamefont {Brennecke}},\ and\ \bibinfo {author} {\bibfnamefont {T.}~\bibnamefont {Esslinger}},\ }\bibfield  {title} {\bibinfo {title} {Dicke quantum phase transition with a superfluid gas in an optical cavity},\ }\href {https://doi.org/10.1038/nature09009} {\bibfield  {journal} {\bibinfo  {journal} {Nature}\ }\textbf {\bibinfo {volume} {464}},\ \bibinfo {pages} {1301} (\bibinfo {year} {2010})}\BibitemShut {NoStop}%
\bibitem [{\citenamefont {Nagy}\ \emph {et~al.}(2010)\citenamefont {Nagy}, \citenamefont {Kónya}, \citenamefont {Szirmai},\ and\ \citenamefont {Domokos}}]{nagy_dicke-model_2010}%
  \BibitemOpen
  \bibfield  {author} {\bibinfo {author} {\bibfnamefont {D.}~\bibnamefont {Nagy}}, \bibinfo {author} {\bibfnamefont {G.}~\bibnamefont {Kónya}}, \bibinfo {author} {\bibfnamefont {G.}~\bibnamefont {Szirmai}},\ and\ \bibinfo {author} {\bibfnamefont {P.}~\bibnamefont {Domokos}},\ }\bibfield  {title} {\bibinfo {title} {Dicke-{Model} {Phase} {Transition} in the {Quantum} {Motion} of a {Bose}-{Einstein} {Condensate} in an {Optical} {Cavity}},\ }\href {https://doi.org/10.1103/PhysRevLett.104.130401} {\bibfield  {journal} {\bibinfo  {journal} {Phys. Rev. Lett.}\ }\textbf {\bibinfo {volume} {104}},\ \bibinfo {pages} {130401} (\bibinfo {year} {2010})}\BibitemShut {NoStop}%
\bibitem [{\citenamefont {Léonard}\ \emph {et~al.}(2017)\citenamefont {Léonard}, \citenamefont {Morales}, \citenamefont {Zupancic}, \citenamefont {Esslinger},\ and\ \citenamefont {Donner}}]{leonard_supersolid_2017}%
  \BibitemOpen
  \bibfield  {author} {\bibinfo {author} {\bibfnamefont {J.}~\bibnamefont {Léonard}}, \bibinfo {author} {\bibfnamefont {A.}~\bibnamefont {Morales}}, \bibinfo {author} {\bibfnamefont {P.}~\bibnamefont {Zupancic}}, \bibinfo {author} {\bibfnamefont {T.}~\bibnamefont {Esslinger}},\ and\ \bibinfo {author} {\bibfnamefont {T.}~\bibnamefont {Donner}},\ }\bibfield  {title} {\bibinfo {title} {Supersolid formation in a quantum gas breaking a continuous translational symmetry},\ }\href {https://doi.org/10.1038/nature21067} {\bibfield  {journal} {\bibinfo  {journal} {Nature}\ }\textbf {\bibinfo {volume} {543}},\ \bibinfo {pages} {87} (\bibinfo {year} {2017})}\BibitemShut {NoStop}%
\bibitem [{\citenamefont {Kollár}\ \emph {et~al.}(2017)\citenamefont {Kollár}, \citenamefont {Papageorge}, \citenamefont {Vaidya}, \citenamefont {Guo}, \citenamefont {Keeling},\ and\ \citenamefont {Lev}}]{kollar_supermode-density-wave-polariton_2017}%
  \BibitemOpen
  \bibfield  {author} {\bibinfo {author} {\bibfnamefont {A.~J.}\ \bibnamefont {Kollár}}, \bibinfo {author} {\bibfnamefont {A.~T.}\ \bibnamefont {Papageorge}}, \bibinfo {author} {\bibfnamefont {V.~D.}\ \bibnamefont {Vaidya}}, \bibinfo {author} {\bibfnamefont {Y.}~\bibnamefont {Guo}}, \bibinfo {author} {\bibfnamefont {J.}~\bibnamefont {Keeling}},\ and\ \bibinfo {author} {\bibfnamefont {B.~L.}\ \bibnamefont {Lev}},\ }\bibfield  {title} {\bibinfo {title} {Supermode-density-wave-polariton condensation with a {Bose}–{Einstein} condensate in a multimode cavity},\ }\href {https://doi.org/10.1038/ncomms14386} {\bibfield  {journal} {\bibinfo  {journal} {Nat. Commun.}\ }\textbf {\bibinfo {volume} {8}},\ \bibinfo {pages} {14386} (\bibinfo {year} {2017})}\BibitemShut {NoStop}%
\bibitem [{\citenamefont {Kroeze}\ \emph {et~al.}(2018)\citenamefont {Kroeze}, \citenamefont {Guo}, \citenamefont {Vaidya}, \citenamefont {Keeling},\ and\ \citenamefont {Lev}}]{kroeze_spinor_2018}%
  \BibitemOpen
  \bibfield  {author} {\bibinfo {author} {\bibfnamefont {R.~M.}\ \bibnamefont {Kroeze}}, \bibinfo {author} {\bibfnamefont {Y.}~\bibnamefont {Guo}}, \bibinfo {author} {\bibfnamefont {V.~D.}\ \bibnamefont {Vaidya}}, \bibinfo {author} {\bibfnamefont {J.}~\bibnamefont {Keeling}},\ and\ \bibinfo {author} {\bibfnamefont {B.~L.}\ \bibnamefont {Lev}},\ }\bibfield  {title} {\bibinfo {title} {Spinor {Self}-{Ordering} of a {Quantum} {Gas} in a {Cavity}},\ }\href {https://doi.org/10.1103/PhysRevLett.121.163601} {\bibfield  {journal} {\bibinfo  {journal} {Phys. Rev. Lett.}\ }\textbf {\bibinfo {volume} {121}},\ \bibinfo {pages} {163601} (\bibinfo {year} {2018})}\BibitemShut {NoStop}%
\bibitem [{\citenamefont {Keeling}\ \emph {et~al.}(2014)\citenamefont {Keeling}, \citenamefont {Bhaseen},\ and\ \citenamefont {Simons}}]{keeling_fermionic_2014}%
  \BibitemOpen
  \bibfield  {author} {\bibinfo {author} {\bibfnamefont {J.}~\bibnamefont {Keeling}}, \bibinfo {author} {\bibfnamefont {M.}~\bibnamefont {Bhaseen}},\ and\ \bibinfo {author} {\bibfnamefont {B.}~\bibnamefont {Simons}},\ }\bibfield  {title} {\bibinfo {title} {Fermionic {Superradiance} in a {Transversely} {Pumped} {Optical} {Cavity}},\ }\href {https://doi.org/10.1103/PhysRevLett.112.143002} {\bibfield  {journal} {\bibinfo  {journal} {Phys. Rev. Lett.}\ }\textbf {\bibinfo {volume} {112}},\ \bibinfo {pages} {143002} (\bibinfo {year} {2014})}\BibitemShut {NoStop}%
\bibitem [{\citenamefont {Piazza}\ and\ \citenamefont {Strack}(2014)}]{piazza_umklapp_2014}%
  \BibitemOpen
  \bibfield  {author} {\bibinfo {author} {\bibfnamefont {F.}~\bibnamefont {Piazza}}\ and\ \bibinfo {author} {\bibfnamefont {P.}~\bibnamefont {Strack}},\ }\bibfield  {title} {\bibinfo {title} {Umklapp {Superradiance} with a {Collisionless} {Quantum} {Degenerate} {Fermi} {Gas}},\ }\href {https://doi.org/10.1103/PhysRevLett.112.143003} {\bibfield  {journal} {\bibinfo  {journal} {Phys. Rev. Lett.}\ }\textbf {\bibinfo {volume} {112}},\ \bibinfo {pages} {143003} (\bibinfo {year} {2014})}\BibitemShut {NoStop}%
\bibitem [{\citenamefont {Chen}\ \emph {et~al.}(2014)\citenamefont {Chen}, \citenamefont {Yu},\ and\ \citenamefont {Zhai}}]{chen_superradiance_2014}%
  \BibitemOpen
  \bibfield  {author} {\bibinfo {author} {\bibfnamefont {Y.}~\bibnamefont {Chen}}, \bibinfo {author} {\bibfnamefont {Z.}~\bibnamefont {Yu}},\ and\ \bibinfo {author} {\bibfnamefont {H.}~\bibnamefont {Zhai}},\ }\bibfield  {title} {\bibinfo {title} {Superradiance of {Degenerate} {Fermi} {Gases} in a {Cavity}},\ }\href {https://doi.org/10.1103/PhysRevLett.112.143004} {\bibfield  {journal} {\bibinfo  {journal} {Phys. Rev. Lett.}\ }\textbf {\bibinfo {volume} {112}},\ \bibinfo {pages} {143004} (\bibinfo {year} {2014})}\BibitemShut {NoStop}%
\bibitem [{\citenamefont {Zhang}\ \emph {et~al.}(2021)\citenamefont {Zhang}, \citenamefont {Chen}, \citenamefont {Wu}, \citenamefont {Wang}, \citenamefont {Fan}, \citenamefont {Deng},\ and\ \citenamefont {Wu}}]{zhang_observation_2021}%
  \BibitemOpen
  \bibfield  {author} {\bibinfo {author} {\bibfnamefont {X.}~\bibnamefont {Zhang}}, \bibinfo {author} {\bibfnamefont {Y.}~\bibnamefont {Chen}}, \bibinfo {author} {\bibfnamefont {Z.}~\bibnamefont {Wu}}, \bibinfo {author} {\bibfnamefont {J.}~\bibnamefont {Wang}}, \bibinfo {author} {\bibfnamefont {J.}~\bibnamefont {Fan}}, \bibinfo {author} {\bibfnamefont {S.}~\bibnamefont {Deng}},\ and\ \bibinfo {author} {\bibfnamefont {H.}~\bibnamefont {Wu}},\ }\bibfield  {title} {\bibinfo {title} {Observation of a superradiant quantum phase transition in an intracavity degenerate {Fermi} gas},\ }\href {https://doi.org/10.1126/science.abd4385} {\bibfield  {journal} {\bibinfo  {journal} {Science}\ }\textbf {\bibinfo {volume} {373}},\ \bibinfo {pages} {1359} (\bibinfo {year} {2021})}\BibitemShut {NoStop}%
\bibitem [{\citenamefont {Helson}\ \emph {et~al.}(2023)\citenamefont {Helson}, \citenamefont {Zwettler}, \citenamefont {Mivehvar}, \citenamefont {Colella}, \citenamefont {Roux}, \citenamefont {Konishi}, \citenamefont {Ritsch},\ and\ \citenamefont {Brantut}}]{helson_density-wave_2023}%
  \BibitemOpen
  \bibfield  {author} {\bibinfo {author} {\bibfnamefont {V.}~\bibnamefont {Helson}}, \bibinfo {author} {\bibfnamefont {T.}~\bibnamefont {Zwettler}}, \bibinfo {author} {\bibfnamefont {F.}~\bibnamefont {Mivehvar}}, \bibinfo {author} {\bibfnamefont {E.}~\bibnamefont {Colella}}, \bibinfo {author} {\bibfnamefont {K.}~\bibnamefont {Roux}}, \bibinfo {author} {\bibfnamefont {H.}~\bibnamefont {Konishi}}, \bibinfo {author} {\bibfnamefont {H.}~\bibnamefont {Ritsch}},\ and\ \bibinfo {author} {\bibfnamefont {J.-P.}\ \bibnamefont {Brantut}},\ }\bibfield  {title} {\bibinfo {title} {Density-wave ordering in a unitary {Fermi} gas with photon-mediated interactions},\ }\href {https://doi.org/10.1038/s41586-023-06018-3} {\bibfield  {journal} {\bibinfo  {journal} {Nature}\ }\textbf {\bibinfo {volume} {618}},\ \bibinfo {pages} {716} (\bibinfo {year} {2023})}\BibitemShut {NoStop}%
\bibitem [{\citenamefont {Zwettler}\ \emph {et~al.}(2025{\natexlab{a}})\citenamefont {Zwettler}, \citenamefont {Marijanovic}, \citenamefont {Bühler}, \citenamefont {Chattopadhyay}, \citenamefont {Del~Pace}, \citenamefont {Skolc}, \citenamefont {Helson}, \citenamefont {Uchino}, \citenamefont {Demler},\ and\ \citenamefont {Brantut}}]{zwettler_cavity-mediated_2025}%
  \BibitemOpen
  \bibfield  {author} {\bibinfo {author} {\bibfnamefont {T.}~\bibnamefont {Zwettler}}, \bibinfo {author} {\bibfnamefont {F.}~\bibnamefont {Marijanovic}}, \bibinfo {author} {\bibfnamefont {T.}~\bibnamefont {Bühler}}, \bibinfo {author} {\bibfnamefont {S.}~\bibnamefont {Chattopadhyay}}, \bibinfo {author} {\bibfnamefont {G.}~\bibnamefont {Del~Pace}}, \bibinfo {author} {\bibfnamefont {L.}~\bibnamefont {Skolc}}, \bibinfo {author} {\bibfnamefont {V.}~\bibnamefont {Helson}}, \bibinfo {author} {\bibfnamefont {S.}~\bibnamefont {Uchino}}, \bibinfo {author} {\bibfnamefont {E.}~\bibnamefont {Demler}},\ and\ \bibinfo {author} {\bibfnamefont {J.-P.}\ \bibnamefont {Brantut}},\ }\bibfield  {title} {\bibinfo {title} {Cavity-mediated charge and pair-density waves in a unitary {Fermi} gas},\ }\href {https://doi.org/10.1038/s41467-025-67184-8} {\bibfield  {journal} {\bibinfo  {journal} {Nat. Commun.}\ }\textbf {\bibinfo {volume} {17}},\ \bibinfo {pages} {496} (\bibinfo {year} {2025}{\natexlab{a}})}\BibitemShut {NoStop}%
\bibitem [{\citenamefont {Bühler}\ \emph {et~al.}(2026)\citenamefont {Bühler}, \citenamefont {Fabre}, \citenamefont {Bolognini}, \citenamefont {Xue}, \citenamefont {Zwettler}, \citenamefont {Del~Pace},\ and\ \citenamefont {Brantut}}]{buhler_microscopy_2026}%
  \BibitemOpen
  \bibfield  {author} {\bibinfo {author} {\bibfnamefont {T.}~\bibnamefont {Bühler}}, \bibinfo {author} {\bibfnamefont {A.}~\bibnamefont {Fabre}}, \bibinfo {author} {\bibfnamefont {G.}~\bibnamefont {Bolognini}}, \bibinfo {author} {\bibfnamefont {Z.}~\bibnamefont {Xue}}, \bibinfo {author} {\bibfnamefont {T.}~\bibnamefont {Zwettler}}, \bibinfo {author} {\bibfnamefont {G.}~\bibnamefont {Del~Pace}},\ and\ \bibinfo {author} {\bibfnamefont {J.-P.}\ \bibnamefont {Brantut}},\ }\bibfield  {title} {\bibinfo {title} {Microscopy of {Cavity}-{Induced} {Density}-{Wave} {Ordering} in {Ultracold} {Gases}},\ }\href {https://doi.org/10.1103/h3zm-rnnx} {\bibfield  {journal} {\bibinfo  {journal} {Phys. Rev. Lett.}\ }\textbf {\bibinfo {volume} {136}},\ \bibinfo {pages} {143401} (\bibinfo {year} {2026})}\BibitemShut {NoStop}%
\bibitem [{\citenamefont {Orsi}\ \emph {et~al.}(2026)\citenamefont {Orsi}, \citenamefont {Fedotova}, \citenamefont {Bhatt}, \citenamefont {Eichenberger}, \citenamefont {Dubois},\ and\ \citenamefont {Brantut}}]{orsi2026fermipressureassistedcavitysuperradiancemesoscopic}%
  \BibitemOpen
  \bibfield  {author} {\bibinfo {author} {\bibfnamefont {F.}~\bibnamefont {Orsi}}, \bibinfo {author} {\bibfnamefont {E.}~\bibnamefont {Fedotova}}, \bibinfo {author} {\bibfnamefont {R.~P.}\ \bibnamefont {Bhatt}}, \bibinfo {author} {\bibfnamefont {M.}~\bibnamefont {Eichenberger}}, \bibinfo {author} {\bibfnamefont {L.}~\bibnamefont {Dubois}},\ and\ \bibinfo {author} {\bibfnamefont {J.-P.}\ \bibnamefont {Brantut}},\ }\href {https://arxiv.org/abs/2603.08691} {\bibinfo {title} {Fermi-pressure-assisted cavity superradiance in a mesoscopic fermi gas}} (\bibinfo {year} {2026}),\ \Eprint {https://arxiv.org/abs/2603.08691} {arXiv:2603.08691 [cond-mat.quant-gas]} \BibitemShut {NoStop}%
\bibitem [{\citenamefont {Clément}\ \emph {et~al.}(2005)\citenamefont {Clément}, \citenamefont {Varón}, \citenamefont {Hugbart}, \citenamefont {Retter}, \citenamefont {Bouyer}, \citenamefont {Sanchez-Palencia}, \citenamefont {Gangardt}, \citenamefont {Shlyapnikov},\ and\ \citenamefont {Aspect}}]{clement_suppression_2005}%
  \BibitemOpen
  \bibfield  {author} {\bibinfo {author} {\bibfnamefont {D.}~\bibnamefont {Clément}}, \bibinfo {author} {\bibfnamefont {A.~F.}\ \bibnamefont {Varón}}, \bibinfo {author} {\bibfnamefont {M.}~\bibnamefont {Hugbart}}, \bibinfo {author} {\bibfnamefont {J.~A.}\ \bibnamefont {Retter}}, \bibinfo {author} {\bibfnamefont {P.}~\bibnamefont {Bouyer}}, \bibinfo {author} {\bibfnamefont {L.}~\bibnamefont {Sanchez-Palencia}}, \bibinfo {author} {\bibfnamefont {D.~M.}\ \bibnamefont {Gangardt}}, \bibinfo {author} {\bibfnamefont {G.~V.}\ \bibnamefont {Shlyapnikov}},\ and\ \bibinfo {author} {\bibfnamefont {A.}~\bibnamefont {Aspect}},\ }\bibfield  {title} {\bibinfo {title} {Suppression of {Transport} of an {Interacting} {Elongated} {Bose}-{Einstein} {Condensate} in a {Random} {Potential}},\ }\href {https://doi.org/10.1103/PhysRevLett.95.170409} {\bibfield  {journal} {\bibinfo  {journal} {Phys. Rev. Lett.}\ }\textbf {\bibinfo {volume} {95}},\ \bibinfo {pages} {170409} (\bibinfo {year} {2005})}\BibitemShut {NoStop}%
\bibitem [{\citenamefont {Billy}\ \emph {et~al.}(2008)\citenamefont {Billy}, \citenamefont {Josse}, \citenamefont {Zuo}, \citenamefont {Bernard}, \citenamefont {Hambrecht}, \citenamefont {Lugan}, \citenamefont {Clément}, \citenamefont {Sanchez-Palencia}, \citenamefont {Bouyer},\ and\ \citenamefont {Aspect}}]{billy_direct_2008}%
  \BibitemOpen
  \bibfield  {author} {\bibinfo {author} {\bibfnamefont {J.}~\bibnamefont {Billy}}, \bibinfo {author} {\bibfnamefont {V.}~\bibnamefont {Josse}}, \bibinfo {author} {\bibfnamefont {Z.}~\bibnamefont {Zuo}}, \bibinfo {author} {\bibfnamefont {A.}~\bibnamefont {Bernard}}, \bibinfo {author} {\bibfnamefont {B.}~\bibnamefont {Hambrecht}}, \bibinfo {author} {\bibfnamefont {P.}~\bibnamefont {Lugan}}, \bibinfo {author} {\bibfnamefont {D.}~\bibnamefont {Clément}}, \bibinfo {author} {\bibfnamefont {L.}~\bibnamefont {Sanchez-Palencia}}, \bibinfo {author} {\bibfnamefont {P.}~\bibnamefont {Bouyer}},\ and\ \bibinfo {author} {\bibfnamefont {A.}~\bibnamefont {Aspect}},\ }\bibfield  {title} {\bibinfo {title} {Direct observation of {Anderson} localization of matter waves in a controlled disorder},\ }\href {https://doi.org/10.1038/nature07000} {\bibfield  {journal} {\bibinfo  {journal} {Nature}\ }\textbf {\bibinfo {volume} {453}},\ \bibinfo {pages} {891} (\bibinfo {year} {2008})}\BibitemShut {NoStop}%
\bibitem [{\citenamefont {Roati}\ \emph {et~al.}(2008)\citenamefont {Roati}, \citenamefont {D’Errico}, \citenamefont {Fallani}, \citenamefont {Fattori}, \citenamefont {Fort}, \citenamefont {Zaccanti}, \citenamefont {Modugno}, \citenamefont {Modugno},\ and\ \citenamefont {Inguscio}}]{roati_anderson_2008}%
  \BibitemOpen
  \bibfield  {author} {\bibinfo {author} {\bibfnamefont {G.}~\bibnamefont {Roati}}, \bibinfo {author} {\bibfnamefont {C.}~\bibnamefont {D’Errico}}, \bibinfo {author} {\bibfnamefont {L.}~\bibnamefont {Fallani}}, \bibinfo {author} {\bibfnamefont {M.}~\bibnamefont {Fattori}}, \bibinfo {author} {\bibfnamefont {C.}~\bibnamefont {Fort}}, \bibinfo {author} {\bibfnamefont {M.}~\bibnamefont {Zaccanti}}, \bibinfo {author} {\bibfnamefont {G.}~\bibnamefont {Modugno}}, \bibinfo {author} {\bibfnamefont {M.}~\bibnamefont {Modugno}},\ and\ \bibinfo {author} {\bibfnamefont {M.}~\bibnamefont {Inguscio}},\ }\bibfield  {title} {\bibinfo {title} {Anderson localization of a non-interacting {Bose}–{Einstein} condensate},\ }\href {https://doi.org/10.1038/nature07071} {\bibfield  {journal} {\bibinfo  {journal} {Nature}\ }\textbf {\bibinfo {volume} {453}},\ \bibinfo {pages} {895} (\bibinfo {year} {2008})}\BibitemShut {NoStop}%
\bibitem [{\citenamefont {Kondov}\ \emph {et~al.}(2011)\citenamefont {Kondov}, \citenamefont {McGehee}, \citenamefont {Zirbel},\ and\ \citenamefont {DeMarco}}]{kondov_three-dimensional_2011}%
  \BibitemOpen
  \bibfield  {author} {\bibinfo {author} {\bibfnamefont {S.~S.}\ \bibnamefont {Kondov}}, \bibinfo {author} {\bibfnamefont {W.~R.}\ \bibnamefont {McGehee}}, \bibinfo {author} {\bibfnamefont {J.~J.}\ \bibnamefont {Zirbel}},\ and\ \bibinfo {author} {\bibfnamefont {B.}~\bibnamefont {DeMarco}},\ }\bibfield  {title} {\bibinfo {title} {Three-{Dimensional} {Anderson} {Localization} of {Ultracold} {Matter}},\ }\href {https://doi.org/10.1126/science.1209019} {\bibfield  {journal} {\bibinfo  {journal} {Science}\ }\textbf {\bibinfo {volume} {334}},\ \bibinfo {pages} {66} (\bibinfo {year} {2011})}\BibitemShut {NoStop}%
\bibitem [{\citenamefont {Jendrzejewski}\ \emph {et~al.}(2012)\citenamefont {Jendrzejewski}, \citenamefont {Bernard}, \citenamefont {Müller}, \citenamefont {Cheinet}, \citenamefont {Josse}, \citenamefont {Piraud}, \citenamefont {Pezzé}, \citenamefont {Sanchez-Palencia}, \citenamefont {Aspect},\ and\ \citenamefont {Bouyer}}]{jendrzejewski_three-dimensional_2012}%
  \BibitemOpen
  \bibfield  {author} {\bibinfo {author} {\bibfnamefont {F.}~\bibnamefont {Jendrzejewski}}, \bibinfo {author} {\bibfnamefont {A.}~\bibnamefont {Bernard}}, \bibinfo {author} {\bibfnamefont {K.}~\bibnamefont {Müller}}, \bibinfo {author} {\bibfnamefont {P.}~\bibnamefont {Cheinet}}, \bibinfo {author} {\bibfnamefont {V.}~\bibnamefont {Josse}}, \bibinfo {author} {\bibfnamefont {M.}~\bibnamefont {Piraud}}, \bibinfo {author} {\bibfnamefont {L.}~\bibnamefont {Pezzé}}, \bibinfo {author} {\bibfnamefont {L.}~\bibnamefont {Sanchez-Palencia}}, \bibinfo {author} {\bibfnamefont {A.}~\bibnamefont {Aspect}},\ and\ \bibinfo {author} {\bibfnamefont {P.}~\bibnamefont {Bouyer}},\ }\bibfield  {title} {\bibinfo {title} {Three-dimensional localization of ultracold atoms in an optical disordered potential},\ }\href {https://doi.org/10.1038/nphys2256} {\bibfield  {journal} {\bibinfo  {journal} {Nat. Phys.}\ }\textbf {\bibinfo {volume} {8}},\ \bibinfo {pages} {398} (\bibinfo {year} {2012})}\BibitemShut {NoStop}%
\bibitem [{\citenamefont {Schreiber}\ \emph {et~al.}(2015)\citenamefont {Schreiber}, \citenamefont {Hodgman}, \citenamefont {Bordia}, \citenamefont {Lüschen}, \citenamefont {Fischer}, \citenamefont {Vosk}, \citenamefont {Altman}, \citenamefont {Schneider},\ and\ \citenamefont {Bloch}}]{schreiber_observation_2015}%
  \BibitemOpen
  \bibfield  {author} {\bibinfo {author} {\bibfnamefont {M.}~\bibnamefont {Schreiber}}, \bibinfo {author} {\bibfnamefont {S.~S.}\ \bibnamefont {Hodgman}}, \bibinfo {author} {\bibfnamefont {P.}~\bibnamefont {Bordia}}, \bibinfo {author} {\bibfnamefont {H.~P.}\ \bibnamefont {Lüschen}}, \bibinfo {author} {\bibfnamefont {M.~H.}\ \bibnamefont {Fischer}}, \bibinfo {author} {\bibfnamefont {R.}~\bibnamefont {Vosk}}, \bibinfo {author} {\bibfnamefont {E.}~\bibnamefont {Altman}}, \bibinfo {author} {\bibfnamefont {U.}~\bibnamefont {Schneider}},\ and\ \bibinfo {author} {\bibfnamefont {I.}~\bibnamefont {Bloch}},\ }\bibfield  {title} {\bibinfo {title} {Observation of many-body localization of interacting fermions in a quasirandom optical lattice},\ }\href {https://doi.org/10.1126/science.aaa7432} {\bibfield  {journal} {\bibinfo  {journal} {Science}\ }\textbf {\bibinfo {volume} {349}},\ \bibinfo {pages} {842} (\bibinfo {year} {2015})}\BibitemShut {NoStop}%
\bibitem [{\citenamefont {Choi}\ \emph {et~al.}(2016)\citenamefont {Choi}, \citenamefont {Hild}, \citenamefont {Zeiher}, \citenamefont {Schauß}, \citenamefont {Rubio-Abadal}, \citenamefont {Yefsah}, \citenamefont {Khemani}, \citenamefont {Huse}, \citenamefont {Bloch},\ and\ \citenamefont {Gross}}]{choi_exploring_2016}%
  \BibitemOpen
  \bibfield  {author} {\bibinfo {author} {\bibfnamefont {J.-y.}\ \bibnamefont {Choi}}, \bibinfo {author} {\bibfnamefont {S.}~\bibnamefont {Hild}}, \bibinfo {author} {\bibfnamefont {J.}~\bibnamefont {Zeiher}}, \bibinfo {author} {\bibfnamefont {P.}~\bibnamefont {Schauß}}, \bibinfo {author} {\bibfnamefont {A.}~\bibnamefont {Rubio-Abadal}}, \bibinfo {author} {\bibfnamefont {T.}~\bibnamefont {Yefsah}}, \bibinfo {author} {\bibfnamefont {V.}~\bibnamefont {Khemani}}, \bibinfo {author} {\bibfnamefont {D.~A.}\ \bibnamefont {Huse}}, \bibinfo {author} {\bibfnamefont {I.}~\bibnamefont {Bloch}},\ and\ \bibinfo {author} {\bibfnamefont {C.}~\bibnamefont {Gross}},\ }\bibfield  {title} {\bibinfo {title} {Exploring the many-body localization transition in two dimensions},\ }\href {https://doi.org/10.1126/science.aaf8834} {\bibfield  {journal} {\bibinfo  {journal} {Science}\ }\textbf {\bibinfo {volume} {352}},\ \bibinfo {pages} {1547} (\bibinfo {year} {2016})}\BibitemShut {NoStop}%
\bibitem [{\citenamefont {Lüschen}\ \emph {et~al.}(2017)\citenamefont {Lüschen}, \citenamefont {Bordia}, \citenamefont {Scherg}, \citenamefont {Alet}, \citenamefont {Altman}, \citenamefont {Schneider},\ and\ \citenamefont {Bloch}}]{luschen_observation_2017}%
  \BibitemOpen
  \bibfield  {author} {\bibinfo {author} {\bibfnamefont {H.~P.}\ \bibnamefont {Lüschen}}, \bibinfo {author} {\bibfnamefont {P.}~\bibnamefont {Bordia}}, \bibinfo {author} {\bibfnamefont {S.}~\bibnamefont {Scherg}}, \bibinfo {author} {\bibfnamefont {F.}~\bibnamefont {Alet}}, \bibinfo {author} {\bibfnamefont {E.}~\bibnamefont {Altman}}, \bibinfo {author} {\bibfnamefont {U.}~\bibnamefont {Schneider}},\ and\ \bibinfo {author} {\bibfnamefont {I.}~\bibnamefont {Bloch}},\ }\bibfield  {title} {\bibinfo {title} {Observation of {Slow} {Dynamics} near the {Many}-{Body} {Localization} {Transition} in {One}-{Dimensional} {Quasiperiodic} {Systems}},\ }\href {https://doi.org/10.1103/PhysRevLett.119.260401} {\bibfield  {journal} {\bibinfo  {journal} {Phys. Rev. Lett.}\ }\textbf {\bibinfo {volume} {119}},\ \bibinfo {pages} {260401} (\bibinfo {year} {2017})}\BibitemShut {NoStop}%
\bibitem [{\citenamefont {Lukin}\ \emph {et~al.}(2019)\citenamefont {Lukin}, \citenamefont {Rispoli}, \citenamefont {Schittko}, \citenamefont {Tai}, \citenamefont {Kaufman}, \citenamefont {Choi}, \citenamefont {Khemani}, \citenamefont {Léonard},\ and\ \citenamefont {Greiner}}]{lukin_probing_2019}%
  \BibitemOpen
  \bibfield  {author} {\bibinfo {author} {\bibfnamefont {A.}~\bibnamefont {Lukin}}, \bibinfo {author} {\bibfnamefont {M.}~\bibnamefont {Rispoli}}, \bibinfo {author} {\bibfnamefont {R.}~\bibnamefont {Schittko}}, \bibinfo {author} {\bibfnamefont {M.~E.}\ \bibnamefont {Tai}}, \bibinfo {author} {\bibfnamefont {A.~M.}\ \bibnamefont {Kaufman}}, \bibinfo {author} {\bibfnamefont {S.}~\bibnamefont {Choi}}, \bibinfo {author} {\bibfnamefont {V.}~\bibnamefont {Khemani}}, \bibinfo {author} {\bibfnamefont {J.}~\bibnamefont {Léonard}},\ and\ \bibinfo {author} {\bibfnamefont {M.}~\bibnamefont {Greiner}},\ }\bibfield  {title} {\bibinfo {title} {Probing entanglement in a many-body–localized system},\ }\href {https://doi.org/10.1126/science.aau0818} {\bibfield  {journal} {\bibinfo  {journal} {Science}\ }\textbf {\bibinfo {volume} {364}},\ \bibinfo {pages} {256} (\bibinfo {year} {2019})}\BibitemShut {NoStop}%
\bibitem [{\citenamefont {Gopalakrishnan}\ \emph {et~al.}(2011)\citenamefont {Gopalakrishnan}, \citenamefont {Lev},\ and\ \citenamefont {Goldbart}}]{gopalakrishnan_frustration_2011}%
  \BibitemOpen
  \bibfield  {author} {\bibinfo {author} {\bibfnamefont {S.}~\bibnamefont {Gopalakrishnan}}, \bibinfo {author} {\bibfnamefont {B.~L.}\ \bibnamefont {Lev}},\ and\ \bibinfo {author} {\bibfnamefont {P.~M.}\ \bibnamefont {Goldbart}},\ }\bibfield  {title} {\bibinfo {title} {Frustration and {Glassiness} in {Spin} {Models} with {Cavity}-{Mediated} {Interactions}},\ }\href {https://doi.org/10.1103/PhysRevLett.107.277201} {\bibfield  {journal} {\bibinfo  {journal} {Phys. Rev. Lett.}\ }\textbf {\bibinfo {volume} {107}},\ \bibinfo {pages} {277201} (\bibinfo {year} {2011})}\BibitemShut {NoStop}%
\bibitem [{\citenamefont {Marsh}\ \emph {et~al.}(2021)\citenamefont {Marsh}, \citenamefont {Guo}, \citenamefont {Kroeze}, \citenamefont {Gopalakrishnan}, \citenamefont {Ganguli}, \citenamefont {Keeling},\ and\ \citenamefont {Lev}}]{marsh_enhancing_2021}%
  \BibitemOpen
  \bibfield  {author} {\bibinfo {author} {\bibfnamefont {B.~P.}\ \bibnamefont {Marsh}}, \bibinfo {author} {\bibfnamefont {Y.}~\bibnamefont {Guo}}, \bibinfo {author} {\bibfnamefont {R.~M.}\ \bibnamefont {Kroeze}}, \bibinfo {author} {\bibfnamefont {S.}~\bibnamefont {Gopalakrishnan}}, \bibinfo {author} {\bibfnamefont {S.}~\bibnamefont {Ganguli}}, \bibinfo {author} {\bibfnamefont {J.}~\bibnamefont {Keeling}},\ and\ \bibinfo {author} {\bibfnamefont {B.~L.}\ \bibnamefont {Lev}},\ }\bibfield  {title} {\bibinfo {title} {Enhancing {Associative} {Memory} {Recall} and {Storage} {Capacity} {Using} {Confocal} {Cavity} {QED}},\ }\href {https://doi.org/10.1103/PhysRevX.11.021048} {\bibfield  {journal} {\bibinfo  {journal} {Phys. Rev. X}\ }\textbf {\bibinfo {volume} {11}},\ \bibinfo {pages} {021048} (\bibinfo {year} {2021})}\BibitemShut {NoStop}%
\bibitem [{\citenamefont {Marsh}\ \emph {et~al.}(2024)\citenamefont {Marsh}, \citenamefont {Kroeze}, \citenamefont {Ganguli}, \citenamefont {Gopalakrishnan}, \citenamefont {Keeling},\ and\ \citenamefont {Lev}}]{marsh_entanglement_2024}%
  \BibitemOpen
  \bibfield  {author} {\bibinfo {author} {\bibfnamefont {B.~P.}\ \bibnamefont {Marsh}}, \bibinfo {author} {\bibfnamefont {R.~M.}\ \bibnamefont {Kroeze}}, \bibinfo {author} {\bibfnamefont {S.}~\bibnamefont {Ganguli}}, \bibinfo {author} {\bibfnamefont {S.}~\bibnamefont {Gopalakrishnan}}, \bibinfo {author} {\bibfnamefont {J.}~\bibnamefont {Keeling}},\ and\ \bibinfo {author} {\bibfnamefont {B.~L.}\ \bibnamefont {Lev}},\ }\bibfield  {title} {\bibinfo {title} {Entanglement and {Replica} {Symmetry} {Breaking} in a {Driven}-{Dissipative} {Quantum} {Spin} {Glass}},\ }\href {https://doi.org/10.1103/PhysRevX.14.011026} {\bibfield  {journal} {\bibinfo  {journal} {Phys. Rev. X}\ }\textbf {\bibinfo {volume} {14}},\ \bibinfo {pages} {011026} (\bibinfo {year} {2024})}\BibitemShut {NoStop}%
\bibitem [{\citenamefont {Kroeze}\ \emph {et~al.}(2025)\citenamefont {Kroeze}, \citenamefont {Marsh}, \citenamefont {Atri~Schuller}, \citenamefont {Hunt}, \citenamefont {Bourzutschky}, \citenamefont {Winer}, \citenamefont {Gopalakrishnan}, \citenamefont {Keeling},\ and\ \citenamefont {Lev}}]{kroeze_directly_2025}%
  \BibitemOpen
  \bibfield  {author} {\bibinfo {author} {\bibfnamefont {R.~M.}\ \bibnamefont {Kroeze}}, \bibinfo {author} {\bibfnamefont {B.~P.}\ \bibnamefont {Marsh}}, \bibinfo {author} {\bibfnamefont {D.}~\bibnamefont {Atri~Schuller}}, \bibinfo {author} {\bibfnamefont {H.~S.}\ \bibnamefont {Hunt}}, \bibinfo {author} {\bibfnamefont {A.~N.}\ \bibnamefont {Bourzutschky}}, \bibinfo {author} {\bibfnamefont {M.}~\bibnamefont {Winer}}, \bibinfo {author} {\bibfnamefont {S.}~\bibnamefont {Gopalakrishnan}}, \bibinfo {author} {\bibfnamefont {J.}~\bibnamefont {Keeling}},\ and\ \bibinfo {author} {\bibfnamefont {B.~L.}\ \bibnamefont {Lev}},\ }\bibfield  {title} {\bibinfo {title} {Directly observing replica symmetry breaking in a vector quantum-optical spin glass},\ }\href {https://doi.org/10.1126/science.adu7710} {\bibfield  {journal} {\bibinfo  {journal} {Science}\ }\textbf {\bibinfo {volume} {389}},\ \bibinfo {pages} {1122} (\bibinfo {year} {2025})}\BibitemShut {NoStop}%
\bibitem [{\citenamefont {Marsh}\ \emph {et~al.}(2025)\citenamefont {Marsh}, \citenamefont {Schuller}, \citenamefont {Ji}, \citenamefont {Hunt}, \citenamefont {Socolof}, \citenamefont {Bowman}, \citenamefont {Keeling},\ and\ \citenamefont {Lev}}]{marsh_multimode_2025}%
  \BibitemOpen
  \bibfield  {author} {\bibinfo {author} {\bibfnamefont {B.~P.}\ \bibnamefont {Marsh}}, \bibinfo {author} {\bibfnamefont {D.~A.}\ \bibnamefont {Schuller}}, \bibinfo {author} {\bibfnamefont {Y.}~\bibnamefont {Ji}}, \bibinfo {author} {\bibfnamefont {H.~S.}\ \bibnamefont {Hunt}}, \bibinfo {author} {\bibfnamefont {G.~Z.}\ \bibnamefont {Socolof}}, \bibinfo {author} {\bibfnamefont {D.~P.}\ \bibnamefont {Bowman}}, \bibinfo {author} {\bibfnamefont {J.}~\bibnamefont {Keeling}},\ and\ \bibinfo {author} {\bibfnamefont {B.~L.}\ \bibnamefont {Lev}},\ }\bibfield  {title} {\bibinfo {title} {Multimode {Cavity} {QED} {Ising} {Spin} {Glass}},\ }\href {https://doi.org/10.1103/x19r-pzyb} {\bibfield  {journal} {\bibinfo  {journal} {Phys. Rev. Lett.}\ }\textbf {\bibinfo {volume} {135}},\ \bibinfo {pages} {160403} (\bibinfo {year} {2025})}\BibitemShut {NoStop}%
\bibitem [{\citenamefont {Lippe}\ \emph {et~al.}(2021)\citenamefont {Lippe}, \citenamefont {Klas}, \citenamefont {Bender}, \citenamefont {Mischke}, \citenamefont {Niederprüm},\ and\ \citenamefont {Ott}}]{lippe_experimental_2021}%
  \BibitemOpen
  \bibfield  {author} {\bibinfo {author} {\bibfnamefont {C.}~\bibnamefont {Lippe}}, \bibinfo {author} {\bibfnamefont {T.}~\bibnamefont {Klas}}, \bibinfo {author} {\bibfnamefont {J.}~\bibnamefont {Bender}}, \bibinfo {author} {\bibfnamefont {P.}~\bibnamefont {Mischke}}, \bibinfo {author} {\bibfnamefont {T.}~\bibnamefont {Niederprüm}},\ and\ \bibinfo {author} {\bibfnamefont {H.}~\bibnamefont {Ott}},\ }\bibfield  {title} {\bibinfo {title} {Experimental realization of a {3D} random hopping model},\ }\href {https://doi.org/10.1038/s41467-021-27243-2} {\bibfield  {journal} {\bibinfo  {journal} {Nat. Commun.}\ }\textbf {\bibinfo {volume} {12}},\ \bibinfo {pages} {6976} (\bibinfo {year} {2021})}\BibitemShut {NoStop}%
\bibitem [{\citenamefont {Periwal}\ \emph {et~al.}(2021)\citenamefont {Periwal}, \citenamefont {Cooper}, \citenamefont {Kunkel}, \citenamefont {Wienand}, \citenamefont {Davis},\ and\ \citenamefont {Schleier-Smith}}]{periwal_programmable_2021}%
  \BibitemOpen
  \bibfield  {author} {\bibinfo {author} {\bibfnamefont {A.}~\bibnamefont {Periwal}}, \bibinfo {author} {\bibfnamefont {E.~S.}\ \bibnamefont {Cooper}}, \bibinfo {author} {\bibfnamefont {P.}~\bibnamefont {Kunkel}}, \bibinfo {author} {\bibfnamefont {J.~F.}\ \bibnamefont {Wienand}}, \bibinfo {author} {\bibfnamefont {E.~J.}\ \bibnamefont {Davis}},\ and\ \bibinfo {author} {\bibfnamefont {M.}~\bibnamefont {Schleier-Smith}},\ }\bibfield  {title} {\bibinfo {title} {Programmable interactions and emergent geometry in an array of atom clouds},\ }\href {https://doi.org/10.1038/s41586-021-04156-0} {\bibfield  {journal} {\bibinfo  {journal} {Nature}\ }\textbf {\bibinfo {volume} {600}},\ \bibinfo {pages} {630} (\bibinfo {year} {2021})}\BibitemShut {NoStop}%
\bibitem [{\citenamefont {Sauerwein}\ \emph {et~al.}(2023)\citenamefont {Sauerwein}, \citenamefont {Orsi}, \citenamefont {Uhrich}, \citenamefont {Bandyopadhyay}, \citenamefont {Mattiotti}, \citenamefont {Cantat-Moltrecht}, \citenamefont {Pupillo}, \citenamefont {Hauke},\ and\ \citenamefont {Brantut}}]{sauerwein_engineering_2023}%
  \BibitemOpen
  \bibfield  {author} {\bibinfo {author} {\bibfnamefont {N.}~\bibnamefont {Sauerwein}}, \bibinfo {author} {\bibfnamefont {F.}~\bibnamefont {Orsi}}, \bibinfo {author} {\bibfnamefont {P.}~\bibnamefont {Uhrich}}, \bibinfo {author} {\bibfnamefont {S.}~\bibnamefont {Bandyopadhyay}}, \bibinfo {author} {\bibfnamefont {F.}~\bibnamefont {Mattiotti}}, \bibinfo {author} {\bibfnamefont {T.}~\bibnamefont {Cantat-Moltrecht}}, \bibinfo {author} {\bibfnamefont {G.}~\bibnamefont {Pupillo}}, \bibinfo {author} {\bibfnamefont {P.}~\bibnamefont {Hauke}},\ and\ \bibinfo {author} {\bibfnamefont {J.-P.}\ \bibnamefont {Brantut}},\ }\bibfield  {title} {\bibinfo {title} {Engineering random spin models with atoms in a high-finesse cavity},\ }\href {https://doi.org/10.1038/s41567-023-02033-3} {\bibfield  {journal} {\bibinfo  {journal} {Nat. Phys.}\ }\textbf {\bibinfo {volume} {19}},\ \bibinfo {pages} {1128} (\bibinfo {year} {2023})}\BibitemShut {NoStop}%
\bibitem [{\citenamefont {Orsi}\ \emph {et~al.}(2024)\citenamefont {Orsi}, \citenamefont {Sauerwein}, \citenamefont {Bhatt}, \citenamefont {Faltinath}, \citenamefont {Fedotova}, \citenamefont {Reiter}, \citenamefont {Cantat-Moltrecht},\ and\ \citenamefont {Brantut}}]{orsi_cavity_2024}%
  \BibitemOpen
  \bibfield  {author} {\bibinfo {author} {\bibfnamefont {F.}~\bibnamefont {Orsi}}, \bibinfo {author} {\bibfnamefont {N.}~\bibnamefont {Sauerwein}}, \bibinfo {author} {\bibfnamefont {R.~P.}\ \bibnamefont {Bhatt}}, \bibinfo {author} {\bibfnamefont {J.}~\bibnamefont {Faltinath}}, \bibinfo {author} {\bibfnamefont {E.}~\bibnamefont {Fedotova}}, \bibinfo {author} {\bibfnamefont {N.}~\bibnamefont {Reiter}}, \bibinfo {author} {\bibfnamefont {T.}~\bibnamefont {Cantat-Moltrecht}},\ and\ \bibinfo {author} {\bibfnamefont {J.-P.}\ \bibnamefont {Brantut}},\ }\bibfield  {title} {\bibinfo {title} {Cavity {Microscope} for {Micrometer}-{Scale} {Control} of {Atom}-{Photon} {Interactions}},\ }\href {https://doi.org/10.1103/PRXQuantum.5.040333} {\bibfield  {journal} {\bibinfo  {journal} {PRX Quantum}\ }\textbf {\bibinfo {volume} {5}},\ \bibinfo {pages} {040333} (\bibinfo {year} {2024})}\BibitemShut {NoStop}%
\bibitem [{\citenamefont {Uhrich}\ \emph {et~al.}(2023)\citenamefont {Uhrich}, \citenamefont {Bandyopadhyay}, \citenamefont {Sauerwein}, \citenamefont {Sonner}, \citenamefont {Brantut},\ and\ \citenamefont {Hauke}}]{uhrich_cavity_2023}%
  \BibitemOpen
  \bibfield  {author} {\bibinfo {author} {\bibfnamefont {P.}~\bibnamefont {Uhrich}}, \bibinfo {author} {\bibfnamefont {S.}~\bibnamefont {Bandyopadhyay}}, \bibinfo {author} {\bibfnamefont {N.}~\bibnamefont {Sauerwein}}, \bibinfo {author} {\bibfnamefont {J.}~\bibnamefont {Sonner}}, \bibinfo {author} {\bibfnamefont {J.-P.}\ \bibnamefont {Brantut}},\ and\ \bibinfo {author} {\bibfnamefont {P.}~\bibnamefont {Hauke}},\ }\href {https://arxiv.org/abs/2303.11343} {\bibinfo {title} {A cavity quantum electrodynamics implementation of the sachdev--ye--kitaev model}} (\bibinfo {year} {2023}),\ \Eprint {https://arxiv.org/abs/2303.11343} {arXiv:2303.11343 [quant-ph]} \BibitemShut {NoStop}%
\bibitem [{\citenamefont {Baumgartner}\ \emph {et~al.}(2024)\citenamefont {Baumgartner}, \citenamefont {Pelliconi}, \citenamefont {Bandyopadhyay}, \citenamefont {Orsi}, \citenamefont {Sauerwein}, \citenamefont {Hauke}, \citenamefont {Brantut},\ and\ \citenamefont {Sonner}}]{baumgartner_quantum_2024}%
  \BibitemOpen
  \bibfield  {author} {\bibinfo {author} {\bibfnamefont {R.}~\bibnamefont {Baumgartner}}, \bibinfo {author} {\bibfnamefont {P.}~\bibnamefont {Pelliconi}}, \bibinfo {author} {\bibfnamefont {S.}~\bibnamefont {Bandyopadhyay}}, \bibinfo {author} {\bibfnamefont {F.}~\bibnamefont {Orsi}}, \bibinfo {author} {\bibfnamefont {N.}~\bibnamefont {Sauerwein}}, \bibinfo {author} {\bibfnamefont {P.}~\bibnamefont {Hauke}}, \bibinfo {author} {\bibfnamefont {J.-P.}\ \bibnamefont {Brantut}},\ and\ \bibinfo {author} {\bibfnamefont {J.}~\bibnamefont {Sonner}},\ }\href {https://arxiv.org/abs/2411.17802} {\bibinfo {title} {Quantum simulation of the sachdev-ye-kitaev model using time-dependent disorder in optical cavities}} (\bibinfo {year} {2024}),\ \Eprint {https://arxiv.org/abs/2411.17802} {arXiv:2411.17802 [quant-ph]} \BibitemShut {NoStop}%
\bibitem [{\citenamefont {Baumgartner}\ \emph {et~al.}(2025)\citenamefont {Baumgartner}, \citenamefont {Pelliconi}, \citenamefont {Bandyopadhyay}, \citenamefont {Orsi}, \citenamefont {Hauke}, \citenamefont {Brantut},\ and\ \citenamefont {Sonner}}]{baumgartner_quantum_2025}%
  \BibitemOpen
  \bibfield  {author} {\bibinfo {author} {\bibfnamefont {R.~L.}\ \bibnamefont {Baumgartner}}, \bibinfo {author} {\bibfnamefont {P.}~\bibnamefont {Pelliconi}}, \bibinfo {author} {\bibfnamefont {S.}~\bibnamefont {Bandyopadhyay}}, \bibinfo {author} {\bibfnamefont {F.}~\bibnamefont {Orsi}}, \bibinfo {author} {\bibfnamefont {P.}~\bibnamefont {Hauke}}, \bibinfo {author} {\bibfnamefont {J.-P.}\ \bibnamefont {Brantut}},\ and\ \bibinfo {author} {\bibfnamefont {J.}~\bibnamefont {Sonner}},\ }\href {https://arxiv.org/abs/2512.13774} {\bibinfo {title} {Quantum simulation using trotterized disorder hamiltonians in a single-mode optical cavity}} (\bibinfo {year} {2025}),\ \Eprint {https://arxiv.org/abs/2512.13774} {arXiv:2512.13774 [quant-ph]} \BibitemShut {NoStop}%
\bibitem [{\citenamefont {Solis}\ \emph {et~al.}(2026)\citenamefont {Solis}, \citenamefont {Windey}, \citenamefont {Bandyopadhyay}, \citenamefont {Legramandi},\ and\ \citenamefont {Hauke}}]{solis_single-particle_2026}%
  \BibitemOpen
  \bibfield  {author} {\bibinfo {author} {\bibfnamefont {D.~P.}\ \bibnamefont {Solis}}, \bibinfo {author} {\bibfnamefont {A.}~\bibnamefont {Windey}}, \bibinfo {author} {\bibfnamefont {S.}~\bibnamefont {Bandyopadhyay}}, \bibinfo {author} {\bibfnamefont {A.}~\bibnamefont {Legramandi}},\ and\ \bibinfo {author} {\bibfnamefont {P.}~\bibnamefont {Hauke}},\ }\bibfield  {title} {\bibinfo {title} {From single-particle to many-body chaos in the {Yukawa}-{Sachdev}-{Ye}-{Kitaev} model: {Theory} and a cavity-{QED} proposal},\ }\href {https://doi.org/10.1103/wntd-53rd} {\bibfield  {journal} {\bibinfo  {journal} {Phys. Rev. B}\ }\textbf {\bibinfo {volume} {113}},\ \bibinfo {pages} {184121} (\bibinfo {year} {2026})}\BibitemShut {NoStop}%
\bibitem [{\citenamefont {Marijanović}\ \emph {et~al.}(2026)\citenamefont {Marijanović}, \citenamefont {Chattopadhyay}, \citenamefont {Skolc}, \citenamefont {Zwettler}, \citenamefont {Halati}, \citenamefont {Jäger}, \citenamefont {Giamarchi}, \citenamefont {Brantut},\ and\ \citenamefont {Demler}}]{marijanovic_quench_2026}%
  \BibitemOpen
  \bibfield  {author} {\bibinfo {author} {\bibfnamefont {F.}~\bibnamefont {Marijanović}}, \bibinfo {author} {\bibfnamefont {S.}~\bibnamefont {Chattopadhyay}}, \bibinfo {author} {\bibfnamefont {L.}~\bibnamefont {Skolc}}, \bibinfo {author} {\bibfnamefont {T.}~\bibnamefont {Zwettler}}, \bibinfo {author} {\bibfnamefont {C.-M.}\ \bibnamefont {Halati}}, \bibinfo {author} {\bibfnamefont {S.~B.}\ \bibnamefont {Jäger}}, \bibinfo {author} {\bibfnamefont {T.}~\bibnamefont {Giamarchi}}, \bibinfo {author} {\bibfnamefont {J.-P.}\ \bibnamefont {Brantut}},\ and\ \bibinfo {author} {\bibfnamefont {E.}~\bibnamefont {Demler}},\ }\bibfield  {title} {\bibinfo {title} {Quench {Instabilities} of a {Strongly} {Interacting} {Quantum} {Gas} in an {Optical} {Cavity}},\ }\href {https://doi.org/10.1103/w2vx-t1mr} {\bibfield  {journal} {\bibinfo  {journal} {Phys. Rev. Lett.}\ }\textbf {\bibinfo {volume} {136}},\ \bibinfo {pages} {193401} (\bibinfo {year} {2026})}\BibitemShut {NoStop}%
\bibitem [{\citenamefont {Giuliani}\ and\ \citenamefont {Vignale}(2005)}]{giuliani_quantum_2005}%
  \BibitemOpen
  \bibfield  {author} {\bibinfo {author} {\bibfnamefont {G.}~\bibnamefont {Giuliani}}\ and\ \bibinfo {author} {\bibfnamefont {G.}~\bibnamefont {Vignale}},\ }\href {https://doi.org/10.1017/CBO9780511619915} {\emph {\bibinfo {title} {Quantum {Theory} of the {Electron} {Liquid}}}},\ \bibinfo {edition} {1st}\ ed.\ (\bibinfo  {publisher} {Cambridge University Press},\ \bibinfo {year} {2005})\BibitemShut {NoStop}%
\bibitem [{\citenamefont {Helson}\ \emph {et~al.}(2022)\citenamefont {Helson}, \citenamefont {Zwettler}, \citenamefont {Roux}, \citenamefont {Konishi}, \citenamefont {Uchino},\ and\ \citenamefont {Brantut}}]{helson_optomechanical_2022}%
  \BibitemOpen
  \bibfield  {author} {\bibinfo {author} {\bibfnamefont {V.}~\bibnamefont {Helson}}, \bibinfo {author} {\bibfnamefont {T.}~\bibnamefont {Zwettler}}, \bibinfo {author} {\bibfnamefont {K.}~\bibnamefont {Roux}}, \bibinfo {author} {\bibfnamefont {H.}~\bibnamefont {Konishi}}, \bibinfo {author} {\bibfnamefont {S.}~\bibnamefont {Uchino}},\ and\ \bibinfo {author} {\bibfnamefont {J.-P.}\ \bibnamefont {Brantut}},\ }\bibfield  {title} {\bibinfo {title} {Optomechanical response of a strongly interacting {Fermi} gas},\ }\href {https://doi.org/10.1103/PhysRevResearch.4.033199} {\bibfield  {journal} {\bibinfo  {journal} {Phys. Rev. Res.}\ }\textbf {\bibinfo {volume} {4}},\ \bibinfo {pages} {033199} (\bibinfo {year} {2022})}\BibitemShut {NoStop}%
\bibitem [{\citenamefont {Wu}\ \emph {et~al.}(2023)\citenamefont {Wu}, \citenamefont {Fan}, \citenamefont {Zhang}, \citenamefont {Qi},\ and\ \citenamefont {Wu}}]{wu_signatures_2023}%
  \BibitemOpen
  \bibfield  {author} {\bibinfo {author} {\bibfnamefont {Z.}~\bibnamefont {Wu}}, \bibinfo {author} {\bibfnamefont {J.}~\bibnamefont {Fan}}, \bibinfo {author} {\bibfnamefont {X.}~\bibnamefont {Zhang}}, \bibinfo {author} {\bibfnamefont {J.}~\bibnamefont {Qi}},\ and\ \bibinfo {author} {\bibfnamefont {H.}~\bibnamefont {Wu}},\ }\bibfield  {title} {\bibinfo {title} {Signatures of {Prethermalization} in a {Quenched} {Cavity}-{Mediated} {Long}-{Range} {Interacting} {Fermi} {Gas}},\ }\href {https://doi.org/10.1103/PhysRevLett.131.243401} {\bibfield  {journal} {\bibinfo  {journal} {Phys. Rev. Lett.}\ }\textbf {\bibinfo {volume} {131}},\ \bibinfo {pages} {243401} (\bibinfo {year} {2023})}\BibitemShut {NoStop}%
\bibitem [{\citenamefont {Zwettler}\ \emph {et~al.}(2025{\natexlab{b}})\citenamefont {Zwettler}, \citenamefont {Del~Pace}, \citenamefont {Marijanovic}, \citenamefont {Chattopadhyay}, \citenamefont {Bühler}, \citenamefont {Halati}, \citenamefont {Skolc}, \citenamefont {Tolle}, \citenamefont {Helson}, \citenamefont {Bolognini}, \citenamefont {Fabre}, \citenamefont {Uchino}, \citenamefont {Giamarchi}, \citenamefont {Demler},\ and\ \citenamefont {Brantut}}]{zwettler_nonequilibrium_2025}%
  \BibitemOpen
  \bibfield  {author} {\bibinfo {author} {\bibfnamefont {T.}~\bibnamefont {Zwettler}}, \bibinfo {author} {\bibfnamefont {G.}~\bibnamefont {Del~Pace}}, \bibinfo {author} {\bibfnamefont {F.}~\bibnamefont {Marijanovic}}, \bibinfo {author} {\bibfnamefont {S.}~\bibnamefont {Chattopadhyay}}, \bibinfo {author} {\bibfnamefont {T.}~\bibnamefont {Bühler}}, \bibinfo {author} {\bibfnamefont {C.-M.}\ \bibnamefont {Halati}}, \bibinfo {author} {\bibfnamefont {L.}~\bibnamefont {Skolc}}, \bibinfo {author} {\bibfnamefont {L.}~\bibnamefont {Tolle}}, \bibinfo {author} {\bibfnamefont {V.}~\bibnamefont {Helson}}, \bibinfo {author} {\bibfnamefont {G.}~\bibnamefont {Bolognini}}, \bibinfo {author} {\bibfnamefont {A.}~\bibnamefont {Fabre}}, \bibinfo {author} {\bibfnamefont {S.}~\bibnamefont {Uchino}}, \bibinfo {author} {\bibfnamefont {T.}~\bibnamefont {Giamarchi}}, \bibinfo {author} {\bibfnamefont {E.}~\bibnamefont {Demler}},\ and\ \bibinfo {author} {\bibfnamefont {J.-P.}\ \bibnamefont {Brantut}},\ }\bibfield  {title} {\bibinfo
  {title} {Nonequilibrium {Dynamics} of {Long}-{Range} {Interacting} {Fermions}},\ }\href {https://doi.org/10.1103/PhysRevX.15.021089} {\bibfield  {journal} {\bibinfo  {journal} {Phys. Rev. X}\ }\textbf {\bibinfo {volume} {15}},\ \bibinfo {pages} {021089} (\bibinfo {year} {2025}{\natexlab{b}})}\BibitemShut {NoStop}%
\bibitem [{\citenamefont {Giorgini}\ \emph {et~al.}(2008)\citenamefont {Giorgini}, \citenamefont {Pitaevskii},\ and\ \citenamefont {Stringari}}]{giorgini_theory_2008}%
  \BibitemOpen
  \bibfield  {author} {\bibinfo {author} {\bibfnamefont {S.}~\bibnamefont {Giorgini}}, \bibinfo {author} {\bibfnamefont {L.~P.}\ \bibnamefont {Pitaevskii}},\ and\ \bibinfo {author} {\bibfnamefont {S.}~\bibnamefont {Stringari}},\ }\bibfield  {title} {\bibinfo {title} {Theory of ultracold atomic {Fermi} gases},\ }\href {https://doi.org/10.1103/RevModPhys.80.1215} {\bibfield  {journal} {\bibinfo  {journal} {Rev. Mod. Phys.}\ }\textbf {\bibinfo {volume} {80}},\ \bibinfo {pages} {1215} (\bibinfo {year} {2008})}\BibitemShut {NoStop}%
\bibitem [{\citenamefont {Zwerger}(2012)}]{zwerger_bcs-bec_2012}%
  \BibitemOpen
  \bibinfo {editor} {\bibfnamefont {W.}~\bibnamefont {Zwerger}},\ ed.,\ \href {https://doi.org/10.1007/978-3-642-21978-8} {\emph {\bibinfo {title} {The {BCS}-{BEC} {Crossover} and the {Unitary} {Fermi} {Gas}}}},\ \bibinfo {series} {Lecture {Notes} in {Physics}}, Vol.\ \bibinfo {volume} {836}\ (\bibinfo  {publisher} {Springer Berlin Heidelberg},\ \bibinfo {address} {Berlin, Heidelberg},\ \bibinfo {year} {2012})\BibitemShut {NoStop}%
\bibitem [{\citenamefont {Duncan}\ and\ \citenamefont {Kirkpatrick}(2008)}]{speckle_sim}%
  \BibitemOpen
  \bibfield  {author} {\bibinfo {author} {\bibfnamefont {D.~D.}\ \bibnamefont {Duncan}}\ and\ \bibinfo {author} {\bibfnamefont {S.~J.}\ \bibnamefont {Kirkpatrick}},\ }\bibfield  {title} {\bibinfo {title} {{Algorithms for simulation of speckle (laser and otherwise)}},\ }in\ \href {https://doi.org/10.1117/12.760518} {\emph {\bibinfo {booktitle} {Complex Dynamics and Fluctuations in Biomedical Photonics V}}},\ Vol.\ \bibinfo {volume} {6855},\ \bibinfo {editor} {edited by\ \bibinfo {editor} {\bibfnamefont {V.~V.}\ \bibnamefont {Tuchin}}\ and\ \bibinfo {editor} {\bibfnamefont {L.~V.}\ \bibnamefont {Wang}}},\ \bibinfo {organization} {International Society for Optics and Photonics}\ (\bibinfo  {publisher} {SPIE},\ \bibinfo {year} {2008})\ p.\ \bibinfo {pages} {685505}\BibitemShut {NoStop}%
\end{thebibliography}
\end{document}